\documentclass[prd, preprint,nofootinbib,superscriptaddress]{revtex4-1}% twocolumn, preprint
\usepackage{epsf,epsfig,subfigure,graphicx,amsmath,amssymb,hyperref,cancel}

\newcommand{\be}{\begin{equation}}
\newcommand{\ee}{\end{equation}}
\newcommand{\bea}{\begin{eqnarray}}
\newcommand{\eea}{\end{eqnarray}}

\def\beq#1\eeq{\begin{align}#1\end{align}}
\def\beqnn#1\eeq{\begin{align*}#1\end{align*}}
\newcommand{\nn}{\nonumber}
\newcommand{\rpv}{$R$-parity violation}
\newcommand{\rpa}{$R$-parity}
\newcommand{\rpcing}{$R$-parity conserving\ }
\newcommand{\rpving}{$R$-parity violating\ }
\def \mt {\tilde m}
\def \mtr { m_{3/2}}

\def \z {_\zeta}
\def \l {\lambda''}
\newcommand{\fru}[2]{\left( \frac{#1}{\text{#2}}\right)}
\newcommand{\frd}[2]{\left( \frac{\text#1}{{#2}}\right)}
\newcommand{\frud}[2]{\left( \frac{#1}{{#2}}\right)}

\def \mtq { m_{\tilde q}}
\def \mtg { m_{\tilde g}}

\def \tev {\text{ TeV}}
\def \gev {\text{ GeV}}
\def \mev {\text{ MeV}}

\newcommand{\OO}{\mathcal{O}}
\newcommand{\LL}{\mathcal{L}}

\newcommand{\hef}{\text{$^4$He}}

\newcommand{\het}{\text{$^3$He}}
\newcommand{\lis}{\text{$^6$Li}}
\newcommand{\lisv}{\text{$^7$Li}}

\begin{document}

%\begin{frontmatter}
\preprint{RUNHETC-2015-03}

\title{Thermal Goldstino Production with Low Reheating Temperatures}
\author{Angelo Monteux}\email{amonteux@physics.rutgers.edu  }
\author{Chang Sub Shin}\email{changsub@physics.rutgers.edu}
\affiliation{New High Energy Theory Center, Department of Physics and Astronomy,
\\Rutgers University, Piscataway, NJ 08854, USA
}
%\emailAdd{amonteux@physics.rutgers.edu  }
%\emailAdd{changsub@physics.rutgers.edu}

%\begin{abstract}
\begin{abstract}
We discuss thermal production of (pseudo) goldstinos, the Goldstone fermions emerging from (multiple) SUSY breaking sectors, when the reheating temperature is well below the superpartner masses. 
In such a case, the production during matter-dominated era induced by inflaton decay stage is more important than after reheating. Depending on the SUSY breaking scale, goldstinos are produced by freeze-in or freeze-out mechanism via $1\to 2$ decays and inverse decays. We solve the Boltzmann equation for the momentum distribution function of the goldstino.
In the freeze-out case, goldstinos maintain chemical equilibrium far after they are kinetically decoupled from the thermal bath, and consequently goldstinos with different momentum decouple at different temperatures.
As a result their momentum distribution function shows a peculiar shape and the final yield is smaller than if kinetic equilibrium was assumed.
%We discuss the production of Goldstinos at reheating temperatures well below the superpartner masses. First, we consider the natural scenario of a matter-dominated era ending at reheating, and find that goldstinos produced in that era can still be abundant enough to be dark matter at present times. Second, goldstinos maintain chemical equilibrium far after they are kinetically decoupled from the thermal bath: as a result their momentum distribution function assumes a peculiar shape and the final yield is smaller than if kinetic equilibrium was assumed. 
We revisit the cosmological implications in both \rpcing and \rpving supersymmetric scenarios. 
For the former, thermally produced goldstinos can still be abundant enough to be dark matter at present times even 
if the reheating temperature is low, of order $1\gev$.
For the latter, if the  reheating temperature is low, of order $0.1-1\gev$, they are safe from the BBN constraints. 
\end{abstract}

%\begin{keyword}
%\texttt{elsarticle.cls}\sep \LaTeX\sep Elsevier \sep template
%\MSC[2010] 00-01\sep  99-00
%\end{keyword}

%\end{frontmatter}

\maketitle

%%%%%%%%%%%%%%%%%%%%%%%%%%%%%%%%%%
\section{Introduction}
When considering consequences of early universe cosmology, it is customary to assume that all interesting phenomena, such as dark matter production, take place after the inflationary period, and in particular after reheating, which marks the moment when the energy density of the inflaton decay products (radiation) dominates over the inflaton energy density. This is justifiable, as any primordial abundance was inflated away and large amounts of entropy are injected during the inflaton decay stage, further diluting any other particle produced during reheating.

However if the reheating temperature $T_R$, the maximum temperature of the thermal bath in the radiation-dominated era, 
is well below the  mass scale relevant for  the particle production, the abundance produced during the inflaton decay stage could be more important than the one produced after reheating. 
There are several early works to calculate the production of WIMP-like particles and its phenomenological consequences during this period when the reheating temperature is low enough \cite{Chung:1998rq,Giudice:2000ex,Roszkowski:2014lga}. In this case, the inflaton decay stage can be approximated as matter-dominated era with constantly injected radiation, and the thermal production can be calculated independently from the specific inflation model. The higher temperature can be achieved during this era and naive (Boltzmann) exponential suppression fpr the abundance is replaced by power suppression. 
On the other hand, 
in the studies of thermal production of super-weakly interacting particles (SWIMP), e.g. gravitinos and axions, such consideration have been ignored so far, with many works focusing on the production at high $T_R$.\footnote{Ref. \cite{Rychkov:2007uq} 
studied the production of gravitinos taking into account a proper treatment of the reheating process, with $T_R\gg\mt$.
 In \cite{Kim:2008yu}, the axino freeze-in thermal production from the neutralino decays is shortly discussed 
for low reheating temperature after thermal inflation.}

One well motivated  class of SWIMP is a  goldstone- or goldstino-like particle, $\zeta$, whose interaction to the visible sector is suppressed by the symmetry breaking scale, and this scale is much higher than its mass.  This class of particles is particularly 
interesting in the sense that their elastic scattering rate is doubly  suppressed by the symmetry breaking scale compared to their production rate. Therefore kinetic decoupling happens earlier than chemical decoupling. Actually this does not give any 
difference if the production happens at high $T_R$ or during radiation dominated era, because chemical interactions also give a thermal distribution for $\zeta$ as $f_\zeta(p)\propto e^{- p/T}$. % \cite{Giudice:2000ex}. 
 However if $T_R$ is well below the relevant particle's mass that produces $\zeta$, and the production during early matter-dominated era is important, the situation is different. As we will show below, the production rate at low temperature is quite momentum dependent, and $f_\zeta(p)$ is no longer proportional to the thermal distribution. Their inverse decay rate becomes more complicated, so a detailed treatment of Boltzmann equations is needed especially when the symmetry breaking scale is small such as in low-scale gauge mediation. 

%It is important to note that the highest experimentally-probed temperature of the radiation-dominated universe %(in a thermal bath) is about $1-10\mev$, corresponding to the temperatures at which Big Bang Nucleosynthesis (BBN) begins. %(at time $\OO(1)\sec$ scales). On the other hand, there is substantial evidence that the universe went through an inflationary phase, although the energy scale of inflation has not been probed yet. After inflation, the inflaton field has to decay into Standard Model (SM) particles to repopulate an empty universe, and this usually occurs while the field coherent oscillates around its minimum, with the inflaton energy density reproducing the matter-dominated scaling $\rho\sim a^{-3}$. Consequently, there is strong evidence for a transition between a matter-dominated and a radiation-dominated era.

%In general, the reheating temperature $T_R$, defined as the maximum temperature of the thermal bath in the radiation-dominated era, is a parameter depending on the interactions between the inflaton and the SM fields. Without a specific model of inflation and reheating, $T_R$ can be taken as a free parameter, free to vary between $1\mev$ and $10^{16}\gev$.

%On the other hand,  a higher temperature of the subdominant radiation was reached during matter domination. It is then possible that the universe was never reheated above the weak scale in the radiation-dominated era, and the only period in which the temperature was high enough to populate additional sectors was before reheating.
In this article, we study thermal production of SWIMPs at low reheating temperature 
considering all the aspects above. We will focus on the production of the goldstino in supersymmetric theories, 
but our treatment would apply equally to similar particles.
In theories where 
%In the study of many super-weakly interacting particles, e.g. gravitinos and axinos, the produced density is proportional to the reheating temperature, because the interactions are dominated by high-energy contributions, and the particles are created out of equilibrium. In particular, they are produced in the scattering of superpartners, such as gluinos and squarks, but only if the reheating temperature is higher than the superpartner mass scale, which we will denote by $\mt$ (to be precise, production is still active down to temperatures around $\mt/10$, thanks to the tails of the Boltzmann distribution).
%Because a proper treating of the reheating process allows for higher temperatures to be achieved during the matter-dominated era, production of gravitinos (and axinos) can be substantial even with very low reheating temperatures, $T_R\lesssim 1\gev \ll 100\gev$.\footnote{% In a similar approach, the authors of  Refs. \cite{Giudice:2000ex,Roszkowski:2014lga} considered the production of weakly-interacting massive particles (WIMPs) by taking a reheating temperature well below the freeze-out temperature.}
 local supersymmetry (SUSY) is broken spontaneously, the resulting goldstino is incorporated in the spin-1/2 degrees of freedom of the gravitino, which acquires a mass $\mtr=F/\sqrt3 M_P$,  $F$ being the scale of SUSY breaking and $M_P=(8\pi G_N)^{-1/2}=2.4\times 10^{18} \gev$ the reduced Planck mass.
Because the coupling between the goldstino and the visible sector is suppressed by $1/F$, 
their mass and thermal production rate are tightly related. 
%for large SUSY breaking scale as in gravity mediation, their thermal production is quite suppressed unless $T_R$ is higher than the sparticle masses, $\tilde m$.  For small SUSY breaking scale, as in low-scale gauge mediation,  goldstino production is active and its number density could be large. But, at the same time, its mass is also very small as of ${\cal O}(10 eV)$, thus the gravitino contribution to the dark matter energy density, $\Omega_{3/2} h^2$, is negligible.
The relation between the goldstino mass and the interaction strength can be changed  %the gravitino mass is $\mathcal{O}(10\, eV)$ and of limited phenomenological interest: for example, it cannot be a Dark Matter component. The reason is that the gravitino mass $\mtr$ is linked to its coupling $1/F$, and for small $F$, which maximizes the number of produced gravitinos, the gravitino contribution to the dark matter energy density, $\Omega_{3/2} h^2$, is too small. %(being proportional to the gravitino mass)
if there are multiple sectors with independent SUSY breaking interactions: multiple {\it goldstini} \cite{Cheung:2010mc} arise and the scenario deserves more phenomenological interest. In particular, while one goldstino is still eaten by the gravitino, there are uneaten goldstini, whose mass is not unambiguously set: at first, in Ref.~\cite{Cheung:2010mc}, it was shown that they generally acquire a mass $2\mtr$ from supergravity interactions. This was extended in Ref. \cite{Craig:2010yf}, where it was computed that the goldstino mass could vary around $2\mtr$ depending on the SUSY breaking dynamics, and in Ref.~\cite{Argurio:2011hs}, where it was shown that, even in the global SUSY limit, the uneaten goldstinos could receive a large mass (up to $\OO(100)$\gev) if multiple SUSY breaking sectors communicate with the Standard Model (SM) via gauge interactions: in this last case, the SM fields take the role of messenger fields in mediating SUSY breaking between the two sectors, and large masses for the goldstino can be achieved. The main interesting change to the standard prediction is that the mass of the uneaten goldstinos can be parametrically larger than the gravitino mass, and can be taken as a free parameter depending on the specifics of the SUSY breaking dynamics.

%In the following, we will focus on the production of the goldstino, but our treatment would apply equally to axinos and similar particles. 

This article is structured as follows: 
in Section \ref{goldstino}, we recall and discuss in more details the goldstino interactions.
In Section \ref{mdera}, we discuss the early matter-dominated era and study the production of goldstinos when the reheating temperature is well below the superpartner scale, $T_R\lesssim\mt/10$. If the $F$-term is small, the goldstino is in thermal equilibrium with the MSSM sector during the matter-dominated era and the freeze-out temperature is momentum-dependent. This happens because the $2\to2$ scattering that would bring the goldstinos in kinetic equilibrium has already frozen out, and only the high-momentum goldstinos can efficiently inverse-scatter into the MSSM thermal bath.
We numerically solve the Boltzmann equation for the momentum distribution function (instead of the equation for the number density), and find an analytical solution that reproduces well the numerical results. We compute the resulting goldstino number yield $Y\z=n\z/s$, which is reduced with respect to the results that one finds assuming kinetic equilibrium. Otherwise, for large $F$-term, the goldstino is slowly produced via superpartners decays (freeze-in). We also consider non-thermal production from direct inflaton decays.
We continue in Section \ref{dmbbn}, where we consider the late-time implications of the produced goldstinos. If \rpa\, is conserved, they can be cold dark matter for reheating temperatures of order 1\gev, and overclose the universe for larger $T_R$; on the other hand, if \rpa\, is violated, they can decay and will typically interfere with Big Bang Nucleosynthesis (BBN). We derive bounds on the reheating temperature in the range of $0.5-10\gev$ for given RPV couplings and goldstino masses. We present final remarks and conclude in Section \ref{end}.

\section{Goldstini Interactions}\label{goldstino}
If there is a single SUSY breaking sector, then the corresponding massless fermionic degree of freedom, the goldstino, %emerges in the low-energy theory. After promoting SUSY to a local symmetry and the goldstino 
forms the spin-1/2 degrees of freedom of the gravitino, which has a mass $\mtr=F/\sqrt3 M_P$. %For small SUSY breaking scales, as in low-scale gauge mediation, the gravitino mass is $\mathcal{O}(eV)$ and of
% limited phenomenological interest: for example, it cannot be a Dark Matter component. The reason is that the gravitino mass $\mtr$ is linked to its coupling $1/F$, and for small $F$, which maximizes the number of produced gravitinos, the gravitino contribution to the dark matter energy density, $\Omega_{3/2} h^2$, is too small. %(being proportional to the gravitino mass)
The situation becomes more complicated if there are multiple sectors with independent SUSY breaking dynamics \cite{Cheung:2010mc}: then, in the limit in which each sector is decoupled from the others, each enjoys its own SUSY algebra and, if SUSY breaking occurs independently in each sector, multiple massless goldstinos would arise. Introducing gravitational interactions, the multiple SUSY algebras are broken down to a diagonal subgroup, and only a linear combination of those goldstini is eaten by the gravitino. As discussed above, the mass of the uneaten goldstinos is not proportional to the value of the corresponding $F$-terms, and can be taken as a free parameter. For example, consider the case of two SUSY breaking sectors with hierarchical $F$-terms $F_1\gg F_2$. The gravitino mass is
\beq
\mtr=\frac{\sqrt{F_1^2+F_2^2}}{\sqrt3 M_P}\approx \frac{F_1}{\sqrt3 M_P},
\eeq
while the different goldstini interactions are suppressed by the different $F$-terms. If each sector contributes 
SUSY breaking scalar mass squares $\mt^2_{\phi 1,2}$, and gaugino masses $\mt_{\lambda 1,2}$ 
we have the following coupling to a matter multiplet $(\phi^i,\psi^i)$ and gauge multiplet $(\lambda^a, A^a)$:
\beq\label{goldstini}
\LL_{int}\approx&\frac{\mt_{\phi 1}^2}{F_{eff}} \eta \psi^i\phi^\dag_i
- 
\frac{i\mt_{\lambda 1}}{\sqrt{2}F_{eff}} \eta \sigma^{\mu\nu} \lambda^a F^a_{\mu\nu} 
 + \frac{\mt_{\phi 2}^2}{F_\zeta}\zeta \psi^i \phi^\dag_i 
-\frac{i\mt_{\lambda 2}}{\sqrt{2}F_\zeta} \zeta \sigma^{\mu\nu} \lambda^a F^a_{\mu\nu} +  
h.c..
\eeq
Here we have denoted by $\eta$ the longitudinal component of the gravitino (the eaten goldstino) and by $\zeta$ the uneaten goldstino. $F_{eff} = \sqrt{F_1^2 + F_2^2}\approx F_1$, and $F_\zeta\approx F_2$ are the corresponding $F$-terms of $\eta$ and $\zeta$, respectively. %{\bf When the sparticle masses are dominated by $\tilde m_{\phi 2}^2$, and $\tilde m_{\lambda 2}$, 
%When the contributions to the sparticle masses from the two sectors are comparable (or when the second sector dominates)},
When $|\mt_1|\lesssim|\mt_2|$, the interactions of $\zeta$ are enhanced with respect to those of $\eta$, while %at the same time 
the mass of $\zeta$ can be kept as  a free parameter, much 
greater than $F_2/M_P$. 
%In Ref.~\cite{Cheung:2010mc}, $m_\zeta\simeq 2m_{3/2} ={\cal O}(F_1/M_P)$, while in the models of \cite{Argurio:2011hs}, $m_\zeta={\cal O}(g^2\mt  F_1/((16\pi^2)^2 F_2)$).

In the following, we will focus on the uneaten goldstino $\zeta$ and we will simply refer to it as the goldstino. We will denote its $F$-term by $F_\zeta$. It is understood that the spectrum also includes the gravitino, which can be neglected because of its suppressed interactions.

When the temperature falls well below the sparticle mass scale, $\tilde m$, the effective interactions between 
the goldstino and other light particles are useful to estimate (non-resonant) elastic scattering rates. 
After integrating out sparticles, 
two-goldstino interactions are given as \cite{Komargodski:2009rz,Brignole:1996fn,Luty:1998np,Brignole:1997pe,Lee:1998xh,Brignole:2003cm}
\beq\label{elastic}
{\cal L}_{eff} = \frac{1}{F_\zeta^2} \bar\zeta \bar\psi^i \square \zeta\psi_i
- \frac{i}{2F_\zeta^2} \zeta \sigma^\mu\partial_\nu \bar\zeta  F^{a\rho\nu} F^a_{\mu\rho}.
\eeq 
If $R$-parity violating interactions are introduced in the superpotential as $W= \lambda_{ijk} \Phi_i \Phi_j \Phi_k$, 
there are single-goldstino interactions even after integrating out sfermions \cite{Dudas:2011kt},
\beq\label{RPVint}
{\cal L}^{RPV}_{eff} = \sum_{ijk} \frac{1}{\tilde m_{\phi_k}^2 F_\zeta} \lambda_{ijk} \psi_i \psi_j\square (\zeta \psi_k)  + h.c..
\eeq
This interaction is not only important to determine the life-time of the goldstino, but also to calculate thermal production at low temperatures.

%The main cosmological consequences are twofold \cite{Cheung:2010mc}: first, goldstinos are easily overproduced, and an upper bound on the reheating temperature can be set as 
%\beq
%T_R \lesssim100 \gev\frd{1\gev}{\mz}\frud{F_\zeta}{(10^5\tev)^2}^2\,.
%\eeq
%Also, recall $T_R>100\gev$ for superpartners at the TeV scale. Second, the lightest observable superpartner (LOSP) decays to goldstinos are enhanced, and do not interfere with BBN only for a small $F$-term, $\sqrt{F_2}\lesssim (10^5-10^6)\tev$. 

%From this discussion, it would seem that low-scale SUSY breaking, say $F\simeq (100\tev)^2$, is not acceptable, as it would overproduce goldstino dark matter by about 10 orders of magnitude for $T_R\gtrsim100\gev$. In this work, we will show that lower reheating temperatures, of order $1\gev$ can provide the correct goldstino dark matter abundance. In the next sections. we provide two mechanisms to evade this constraint: first, previous calculations did not include the effect of an early matter-dominated period. As we will see below, even for low $T_R$ it is easy to achieve temperatures larger than $\mt$, thus allowing superpartners in the thermal bath. The resulting gravitinos will be diluted by the entropy injection and can emerge at reheating with the correct relic abundance.
%Second, in the presence of \rpv, goldstino production can happen through the scattering $qq\to q\zeta$ via an off-shell squark, even at thermal bath temperatures well below the squark mass, but above $m_\zeta$.

\section{Goldstino Production in Matter-Dominated Era}\label{mdera}

\subsection{Cosmology during Reheating}
\begin{figure}[t]
\begin{center}
\includegraphics[height=0.25\textheight]{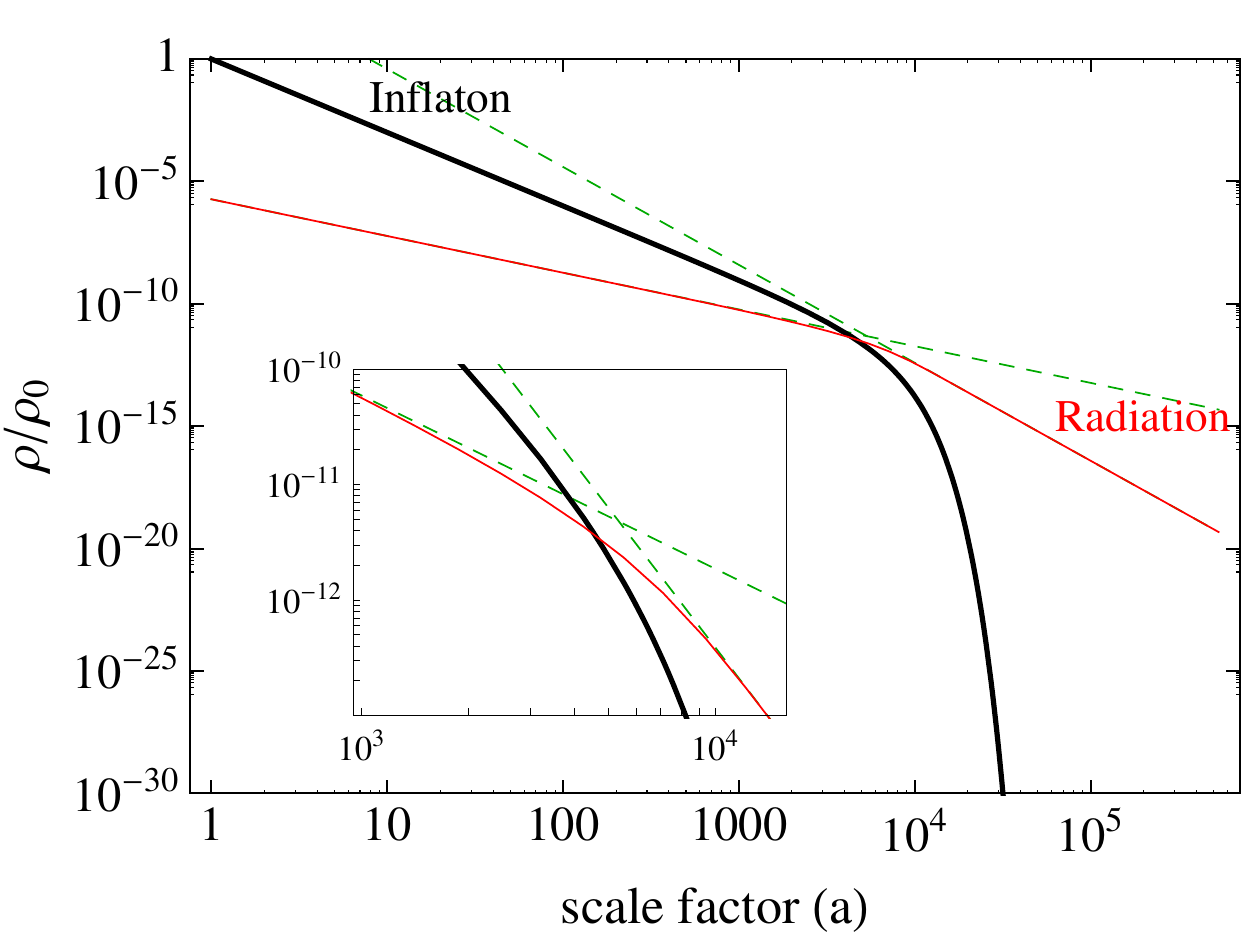}\hfill
\includegraphics[height=0.25\textheight]{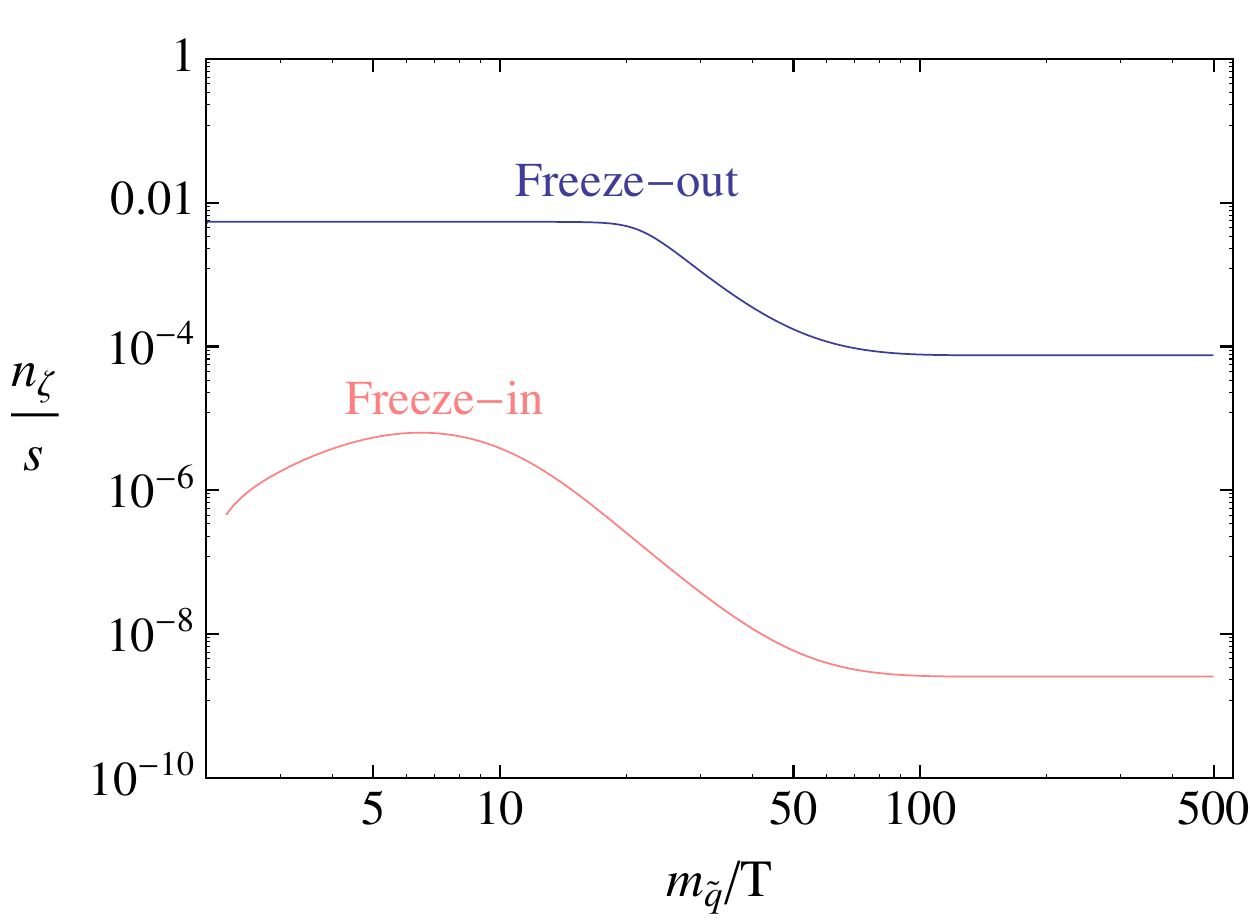}
\caption{Left: schematic behavior of the energy densities of the inflaton and of thermal radiation (MSSM fields) during the matter-dominated era; as $\rho_r\sim T^4$, we have % $T_R= 20\,{\rm GeV}$ and 
$T_{MAX} \gg T_R$.
Right: Evolution of the goldstino yield with fixed reheating temperature $T_R=20\gev$ (corresponding to $a_R=5430$), squark mass $\mtq=1\tev$ and two choices of $F\z=(100\tev)^2,(5000\tev)^2$, corresponding to freeze-out and freeze-in, respectively. 
}
\label{figMDera}
\end{center}
\end{figure}

The reheating temperature is usually referred to as the temperature of the thermal bath when the inflaton  decays,\footnote{%
While we will be referring to the inflaton as the field dominating the energy density of the universe in the early matter-dominated era, there are other cases that reproduce the same scaling of the energy density, $\rho_X\sim a^{-3}$; a typical examples would be the late decay of moduli, which typically result in very low reheating temperatures. In any case, we will refer to the reheating temperature as the maximum temperature achieved in the last radiation-dominated era, the one leading to Big Bang Nucleosynthesis.}
and  it is usually  defined as $H(T_R)=\Gamma_I$, assuming that the radiation energy density, $\rho_r$, dominates the universe. 
In fact, when $t=1/\Gamma_I$, where $\Gamma_I$ is the decay rate of the inflaton, the inflaton energy density is still not-negligible and such a temperature was achieved only during matter domination. Thus entropy injection is still occurring after $T_R$. One more precise way to define $T_R$ 
is to consider the asymptotic behaviors of $\rho_r$ in both matter- and radiation-dominated epochs, as shown in Fig.~\ref{figMDera};
\beq
\rho_r =\frac{\pi^2 g_*(T)}{30} T^4 =\left\{ \begin{array}{ll} 
(\pi^2 g_*(T) T_R^4/30)(a/a_R)^{-3/2}  & {\rm for}\ T \gg T_R,\\ 
(\pi^2 g_*(T) T_R^4/30)(a/a_R)^{-4}  & {\rm for}\ T \ll T_R ,
\end{array}\right.
\eeq
where $g_*(T)$ is the effective number of massless degrees of freedom at the temperature $T$, 
$a$ is the scale factor and $a_R$ is the scale at which two asymptotic lines meet. 
Numerically we get 
% when $H(T_R)=\Gamma_I$:
\beq
%\rho_r=3 H^2 M_P^2=\pi^2g_*T^4/30,\qquad H(T_R)=\Gamma_I \implies
T_R=	0.7\left(\frac{90}{\pi^2 g_*(T_R)} \right)^{1/4}\sqrt{\Gamma_I M_P}\,.
\eeq
%Here $M_P=(8\pi G_N)^{-1/2}=2.4\times 10^{18} \gev$ is the reduced Planck mass.
 This will be our definition of the reheating temperature.
%In the following, we will use a more precise definition of $T_R$, as 

On the other hand, the thermal bath in the matter-dominated era reached a higher temperature,
\beq\label{Tmax}
T_{MAX}=\left(\frac{24g_*(T_R)}{5\pi^2g_*^2(T_{MAX})}\right)^{1/8}%\sqrt{\mu T_R}
\frud{\mu}{T_R}^{1/2}T_R\,,
\eeq
where $\mu=V_I^{1/4}$ is the energy scale of inflation, and slow-roll was assumed for this simplified expression. Because of the continuous entropy injection during the matter-dominated period, the temperature scales as $T/T_{MAX}\propto a^{-3/8}$ instead of the typical scaling of the radiation-dominated era, $T\propto a^{-1}$ (see Fig. \ref{figMDera}). We refer to Ref.~\cite{Giudice:2000ex} for the details of the computations leading to $T_{MAX}$. It is sufficient here to note that $T_{MAX}\gg T_R$ can be much larger than the$\tev$ (sparticle mass) scale, for typical inflation scales $\mu\sim10^8\gev-10^{16}\gev$. The hierarchy of relevant scales can be summarized as 
\beq
m_\psi,\ m_\zeta,\ T_R \ll  \tilde m = {\cal O}({\rm TeV}) \lesssim T_{MAX},
\eeq 
where $\psi$ are SM particles, $m\z$ is the mass of $\zeta$, which is taken as the NLSP by assuming $m_{3/2} < m_\zeta\ll \tilde m$. 
We consider the thermal production of goldstinos at $T\lesssim T_{MAX}$, when $T_R$ is taken low, $T_R = {\cal O}(\rm GeV)$.

The goldstino can be produced (and annihilated) via three different channels: 
%(
i) scattering: $\phi+A_\mu \leftrightarrow \zeta+\psi, \ \lambda+\lambda \leftrightarrow \zeta+ \lambda,\ \ldots$,
%(
ii) (inverse) decay: $\phi \leftrightarrow \zeta+ \psi, \ \lambda \leftrightarrow \zeta +A_\mu,\ \ldots$, and 
%(
iii) RPV scattering: $\psi+\psi\leftrightarrow \zeta+\psi$ (here $\psi,\phi,\lambda,A_\mu$ respectively stand for SM fermions, sfermions,  gauginos, and gauge bosons). %For low $F$-term, these (goldstino-number-changing)  interactions can be in chemical equilibrium at high temperatures, until the interaction rate drops below the Hubble rate and the process freezes out. %It is also possible for the goldstino to elastically scatter off of the thermal bath: when the elastic interaction rate is large enough, the goldstinos would be in kinetic equilibrium with the thermal bath. Because the $\zeta+\psi (A_\mu)\to\zeta+\psi (A_\mu)$ process is suppressed by $F_\zeta^2$, while the single-goldstino production channels are suppressed only by $F\z$, kinetic decoupling takes place before chemical decoupling.\footnote{% This is the opposite behavior than for WIMPs, in which even after chemical decoupling WIMPs elastically scatter off of the thermal bath and remain in kinetic equilibrium untl lower temperatures.} To be precise, the momentum distribution of $\zeta$  is determined not only by elastic scattering but also by chemical interactions. Thus if the production happens at $T \gg \tilde m$, energy-momentum conservation just tells us that the momentum distribution of thermally produced $\zeta$ would be the form of $f_\zeta({\bf p})\propto e^{-p/T}$. Also if the chemical interactions are efficient enough (thus, for small $F_\zeta$), the produced goldstinos will still have a equilibrium distribution function, $f_\zeta({\bf p})= \exp(-p/T)$.    As we will see, goldstinos with different momentum decouple at different temperatures and, in the absence of elastic scattering, do not re-thermalize.

The Boltzmann equation for the number density $n_\zeta$ can be written as
\beq\label{Beq}
\dot n\z+3Hn\z=&g\z\int \frac {d^3 {\bf p}\z}{(2\pi)^3}C[f_\zeta] \\
=&\, \left\langle \sigma_{\phi A_\mu\to\zeta \psi} v\right\rangle_{\phi A_\mu}  n_{\phi } n_{A_\mu}
- \left\langle \sigma_{\zeta \psi \to \phi A_\mu} v\right\rangle_{\zeta \psi } n_{\zeta} n_{\psi}  + \cdots
\,  &{\rm (i)} \nonumber\\
&+\, \left\langle \Gamma_{\phi\to \zeta \psi}\gamma^{-1}\right\rangle_{\phi} n_{\phi} 
- \left\langle \sigma_{\zeta \psi\to \phi}v\right\rangle_{\zeta \psi} n_\zeta n_\psi  + \cdots\,  &{\rm (ii)} \nonumber\\
&+\, \left\langle \sigma_{\psi\psi\to\zeta \psi} v\right\rangle_{\psi\psi} n_\psi n_\psi 
- \left\langle \sigma_{\zeta \psi \to \psi\psi} v\right\rangle_{\zeta \psi} n_\zeta n_\psi, \ & {\rm (iii)} \nn
\eeq
where $\gamma=1/\sqrt{1-v^2}$, 
$n_{\psi_i} \equiv g_{\psi_i} \int d^3{\bf p}_i/(2\pi)^3\,  f_{\psi_i}({\bf p}_i)$ and
\bea 
\left.\langle {\cal O}\rangle_{\psi_i\cdots \psi_j} \equiv \frac{1}{n_{\psi_i}\cdots n_{\psi_j}} 
 \int \frac{g_{\psi_i} d^3{\bf p}_i}{(2\pi)^3}\cdots
 \frac{g_{\psi_j} d^3 {\bf p}_j}{(2\pi)^3} f_{\psi_i} ({\bf p}_i)\cdots f_{\psi_j}({\bf p}_j)\, {\cal O}.\right.
\eea 
Here, $\Gamma_{\phi\to\zeta \psi}$ is defined in the rest frame.
We take $n_\Psi = n_\Psi^{\rm eq}$ for $\Psi = \phi,\, \psi,\, \lambda,\, A_\mu$, 
because they are all in thermal equilibrium for the relevant temperature scale. 
One might rewrite the inverse scattering and decay terms in the the RHS of the Boltzmann equation, and 
Eq.~(\ref{Beq}) becomes 
\beq\label{Beqeq}
\dot n\z+3Hn\z=&\left(\left\langle\sigma_{\phi A_\mu\to\zeta \psi} v\right\rangle_T n_\phi^{\rm eq} n_{A_\mu}^{\rm eq} 
+  \left\langle\Gamma_{\phi\to \zeta \psi}\gamma^{-1}\right\rangle_T n_{\phi}^{\rm eq}
+ \ldots \right)\left( 1-  n_\zeta/n_\zeta^{\rm eq}\right)\nn\\
\equiv&\,
\Gamma_{prod}^\zeta\,  \left(n_\zeta^{\rm eq} -n_\zeta\right),
\eeq where $\langle\cdots \rangle_T$ denotes thermal average. In the treatment of the Boltzmann equation, we have neglected quantum-statistical effects (Pauli-blocking/Bose-enhancement), as we have $f_i\lesssim1$.
For low $F$-term, these (goldstino-number-changing)  interactions can be in chemical equilibrium at high temperatures, until the interaction rate drops below the Hubble rate and the process freezes out. 
The production rate $\Gamma^\zeta_{prod}$ of Eq.~(\ref{Beqeq}) determines the chemical freeze-out (decoupling) temperature, $T_{f.o.}$, defined by $3H(T_{f.o.})=\Gamma^\zeta_{prod}$.
Depending on the value of  $T_{f.o.}$, 
two distinct situations for the goldstino production are possible, displayed in Fig. \ref{figMDera}:
\begin{itemize}
\item freeze-in: for $T_{f.o.}\gg \mt%\frac\mtq{10}
$, goldstino interactions were not in thermal equilibrium when superpartners were abundant; goldstino abundance is gradually increased to a maximum, after which they are diluted. 
\item freeze-out: for $T_{f.o.}\ll\mt% \frac\mtq{10}
$, goldstinos maintain chemical equilibrium with the superpartners until the latter are not abundant.
\end{itemize}

Since the production at  high temperatures $T\gtrsim \tilde m$ is diluted away, we can only focus on the production for $T\lesssim \tilde m$.
At such low temperature, the Boltzmann equation can be much simplified by ignoring the scattering contribution to the production of $\zeta$ \cite{Covi:2002vw,Cheung:2011nn}; for $T_{f.o.}\gg \mt$, the freeze-in production by $\phi\to \zeta+\psi$ ($\lambda\to\zeta+A_\mu$) dominates over the diluted freeze-out contribution. 
One can also neglect the inverse decay term given by $n_\zeta/n_\zeta^{\rm eq}$ in the last expression of the first line of Eq.~(\ref{Beqeq}).

For the freeze-out case, the situation is more subtle. 
It should be noted that the factorization by $(1-n_\zeta/n_\zeta^{\rm eq})$ 
leading to Eq.~\eqref{Beqeq} is only valid if $\zeta$ is in kinetic equilibrium, or at least if $f_\zeta({\bf p}')/f_\zeta({\bf p}) = e^{-(p'-p)/T}$.
It is possible for the goldstino to elastically scatter off of the thermal bath as given by the interactions in Eq.~(\ref{elastic}): if the elastic interaction rate is large enough, the goldstinos would be in kinetic equilibrium with the thermal bath. However, because the $\zeta+\psi (A_\mu)\to\zeta+\psi (A_\mu)$ process is suppressed by $F_\zeta^2$ while the single-goldstino production channels are suppressed only by $F\z$,
kinetic decoupling takes place before chemical decoupling.\footnote{%
This is the opposite behavior than for WIMPs, in which even after chemical decoupling WIMPs elastically scatter off of the thermal bath and remain in kinetic equilibrium until lower temperatures.}
To be precise, the momentum distribution of $\zeta$  is determined not only by elastic scattering but also by chemical interactions. Thus if the production happens at $T \gg \tilde m$, energy-momentum conservation just tells us 
that the momentum distribution of thermally produced $\zeta$ would be the form of $f_\zeta({\bf p})\propto e^{-p/T}$.
Also if the chemical interactions are efficient enough (thus, for small $F_\zeta$), the produced goldstinos will still have a equilibrium distribution function, $f_\zeta({\bf p})= f_\zeta^{\rm eq}= \exp(-p/T)$.    
These arguments are not sufficient to justify the form of equation around the time of decoupling, 
since as we will see, at low $T$ with small $F_\zeta$-term, goldstinos with different momentum decouple at different temperatures and, in the absence of elastic scattering, do not re-thermalize. 
Furthermore, the continuous entropy injection during matter dominated era causes the goldstinos decoupled earlier  to be colder than those decoupled later. 
%If the interaction rates are small, the number density never approaches the equilibrium value $n^{\rm eq}_\zeta$ and one can neglect the $n_\zeta/n_\zeta^{\rm eq}$ term in this last expression. In this case (freeze-in), the assumption of kinetic equilibrium is not important and the final number density is found solving the Boltzmann equation above. Otherwise, 
Therefore it is necessary to solve the non-integrated version of the Boltzmann equation for the distribution function $f\z({\bf p})$:
\beq\label{Boltzmann_dist}
\frac{d f_\zeta}{d t} &=
\frac{\partial f_\zeta}{\partial t}  - H p \frac{\partial f_\zeta}{\partial p}= C[f_\zeta]\,.
\eeq
Substituting $H dt = d\ln a$, for $T\lesssim \tilde m$ this can be rewritten as
\beq\label{BoltzEQ}
\frac{\partial f_\zeta}{\partial \ln a } - \frac{\partial f_\zeta}{\partial \ln p}  =&
\left(1 - \frac{f_\zeta}{e^{-p/T}} \right)
%\times\\\nn
%&\qquad\times
\left(\frac{\Gamma_{\phi\to \zeta \psi} \tilde m_{\phi} T}{ H p^2}\right)\exp\left\{- \frac{p}{T}\left( 1+ \frac{\tilde m_{\phi}^2}{4 p^2}\right)\right\} \nn\\
&\,+ (\phi\to \lambda, \psi\to A_\mu)
\eeq
in the limit of $m\z\to0$. In the Appendix, we provide the Boltzmann equation for non-negligible $m\z\lesssim \mt$. 
%Because the production given by gaugino decays is nothing but a change of mass and degrees of freedoms of sfermion case, here, we will consider one generation of squarks, $\tilde q$,  with mass $\mtq$ of ${\cal O}$(TeV) while the others are all heavy enough. Adding other contributions is straightforward. For simplicity, the number of massless degrees of freedom $g_*(T)$ is taken as a constant, $g_*=85$, in the whole range $T_R\lesssim T\lesssim \mtq$. 
In the following, we will only consider decays from one generation of squarks, $\tilde q$,  with mass $\mtq$ of ${\cal O}$(TeV) while the others are assumed to be heavier. Adding other contributions is straightforward. For example, production by gaugino decays will take the same form with the substitution $\mtq\to\mtg$ and changing the number of sfermions to number of gauginos. For simplicity, the number of massless degrees of freedom $g_*(T)$ is taken constant, $g_*=85$, in the whole range $T_R\lesssim T\lesssim \mtq$. 

The freeze-out temperature for $T_{f.o.}\lesssim \mtq$ 
is calculated from
\beq
&3 H(T_{f.o.})= 1.4\left(\frac{5\pi^2 g_*}{72}\right)^{1/2} \frac{T_{f.o.}^4}{M_P T_R^2}
  \nn\\
=&\sum_{\text{1 gen.}} \left\langle\Gamma_{\tilde q\to \zeta q}\gamma^{-1}\right\rangle_T
n_{\tilde q}^{\rm eq}/n_\zeta^{\rm eq} 
\simeq
\frac{12 \mtq^5}{16\pi F_\zeta^2}\sqrt{\pi}\left(\frac{\mtq}{T_{f.o.}} \right)^{3/2} e^{-\mtq/T_{f.o.}},
\eeq
where $\Gamma_{\tilde q\to \zeta q}=\mtq^5/(16\pi F\z^2)$, and the result is
%Because of the continuous entropy injection, goldstinos produced at early times are diluted and the most important contribution to the late-time abundance comes from the lowest temperature at which the production is active. 
%There are two scales to keep in mind: the freeze-out temperature, defined by $H(T_{f.o.})=\Gamma$, below which goldstinos are not produced efficiently, and the squark mass $\mtq$ (in fact $\mtq/10$ will be more relevant? MAYBE). For the decay process $\tilde q\to q+\zeta$, the decay rate and the goldstino freeze-out temperature are
\beq\label{Tfo}
T_{f.o.}=\frac{\mtq}{21.2+\delta}\,,
\eeq
with $\delta= 5.5\ln\frac{\mtq}{20 T_{f.o.}%^{\rm RPC}
} 
+ \ln\frac{\mtq}{{\rm TeV}} 
+ 2\ln\frac{(100\,{\rm TeV})^2}{F_\zeta}+ \frac{1}{2}\ln\frac{85}{g_*}  
+ 2\ln\frac{T_{\rm R}}{10\, {\rm GeV}}
$.
%Because the superpartner number density decreases for $T<\mt$, the decay process is efficient until temperatures of order $\mtq/10$. Two distinct situations are possible, displayed in Figure \ref{figMDera}:
%\begin{itemize}
%\item freeze-in: for $T_{f.o.}\gg \mtq%\frac\mtq{10}
%$ goldstino interactions were not in thermal equilibrium when superpartners were abundant; goldstino abundance is gradually increased to a maximum, after which they are diluted. 
%\item freeze-out: for $T_{f.o.}\ll\mtq% \frac\mtq{10}
%$, goldstinos maintain chemical equilibrium with the superpartners until the latter are not abundant.
%\end{itemize}

This value is no longer true for large $F$-term. In this case, 
the freeze-out temperature becomes well above $\mtq$, and it is mostly determined by scattering process. 
The freeze-out abundance is quite diluted by entropy production, and 
freeze-in production dominates the goldstino abundance as shown in Fig.~\ref{figMDera}.

In the following, we derive the goldstino yield for each case of freeze-in and out.

\subsection{Freeze-in}
First, we will consider the simpler case of intermediate or high $F_\zeta$, for which goldstinos never reach chemical equilibrium for $T\lesssim \mtq$; their abundance is gradually increased by the thermal decay process (ii) until it is not efficient,
%superpartner number density becomes too low (at around $\mtq/10$)
after which they are diluted during the rest of the matter-dominated era.

For $n_\zeta\ll n_\zeta^{\rm eq}$, the resulting yield at reheating can be computed as:
%\begin{widetext}
\bea\label{nzetaFI}
%\left(
\left(\frac{n_\zeta}{s}\right)_{FI}
%\right)_{\rm RPC} 
&=&\frac{1}{s_R}\int_{t_I}^{t_R} dt\,
\left(\frac{a}{a_R} \right)^3  \sum_{\text{1 gen.}}\langle \Gamma_{\tilde q\to \zeta q } \gamma^{-1}\rangle n_{\tilde q}^{\rm eq}
\nn\\&=&
 \frac{15.6  M_P \sum_{1 gen.}\Gamma_{\tilde q\to\zeta q}}{g_*^{3/2}T_R^5} \int_{T_R}^{T_{f.o.}} 
\frac{d T}{T}  \left(\frac{T_R}{T}\right)^{12} \frac{K_1(\mtq/T)}{K_2(\mtq/T)}\,n_{\tilde q}^{\rm eq}%e^{-\mtq/T}
%f_{\tilde q}^{\rm eq} \left(T\right)
\nn\\
&\simeq %\left\{\begin{array}{ll}   
&2\times 10^{-7}
%\times\\\nn&&\qquad\times
\left(\frac{85}{g_*} \right)^{3/2} 
\left(\frac{T_{\rm R}}{10\, {\rm GeV}}\right)^7 \left(\frac{(500\, {\rm TeV})^2}{F_\zeta}\right)^2 
\left(\frac{\rm TeV}{m_{\tilde q}}\right)^4, 
\eea
where $s_R \equiv (2\pi^2 g_*/45) T_R^3$, and $t_R$ is the time at $a=a_R$. 
$K_{\alpha}$ is the modified Bessel function of the second kind, and $K_1(x)/K_2(x)\simeq 1- 3/(2x)$ for $x\gg 1$.
$t_I$ is the initial time and we took it at the freeze-out of chemical interactions.
The integrand on the second line shows 
 high powers of $(T_R/T)$  caused by 
entropy injection and temperature dependence of the freeze-in production rate, 
and because the Boltzmann suppression factor at low $T$ from $n_{\tilde q}^{\rm eq}$, 
the production rate is most efficient around $T\simeq \mtq/10$. 
The third line of Eq.~(\ref{nzetaFI}) is obtained assuming $T_R\ll \mtq/10\ll T_{f.o.}$. 
Given the expression \eqref{Tfo} for $T_{f.o.}$, this corresponds to $F\z\gg (500\tev)^2$ for $\mtq=1\tev$. 
For lower values of $F\z$, the freeze-out abundance is more important, which will be treated next.
Here we just note that this yield is sizable, and will discuss late-time implications in Section~\ref{dmbbn}.

\subsection{Freeze-out}
For smaller $F\z$, one could solve the same Boltzmann equation, \eqref{Beq}-\eqref{Beqeq}, and find
\beq\label{nzetaeq}
\left(\frac{n_\zeta}{s}\right)_{FO}^{\rm{eq}}=
1.7\times 10^{-6} \left(\frac{85}{g_*}\right)\left(\frac{T_{\rm R}}{10\,{\rm GeV}}\right)^5
\left(\frac{50\,{\rm GeV}}{T_{f.o.}%^{\rm RPC}
}\right)^5.
\eeq
This would be incorrect, because the RHS of the Boltzmann equation \eqref{Beqeq} was found assuming that the goldstinos are in kinetic equilibrium with the rest of the thermal bath. When goldstinos are injected in the bath via decays of non-relativistic particles, their momentum distribution is peaked around $\mtq/2 $. Because the $2\to2$ elastic scattering is frozen out at a higher temperature, %$T_{f.o.}^{\rm{kin}}\sim 10\tev$, 
it does not thermalize the distribution function.  The result is that goldstinos with high momentum easily inverse decay back into superpartners, while goldstinos at low momentum are effectively frozen out.
Therefore one can expect the correct value would be smaller than Eq.~(\ref{nzetaeq}).

\begin{figure}[t]
\begin{center}
\includegraphics[height=0.23\textheight]{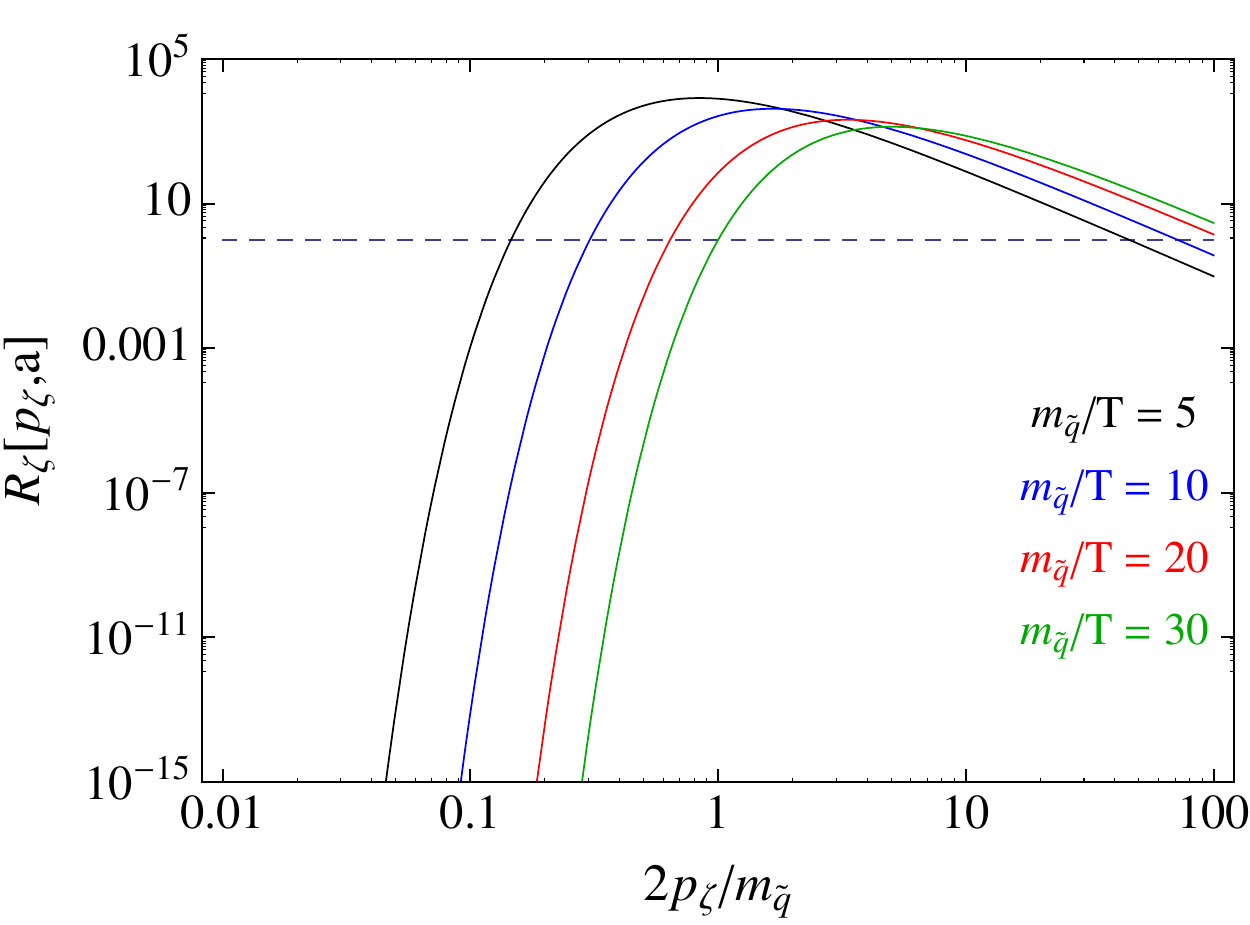}\hfill
\includegraphics[height=0.23\textheight]{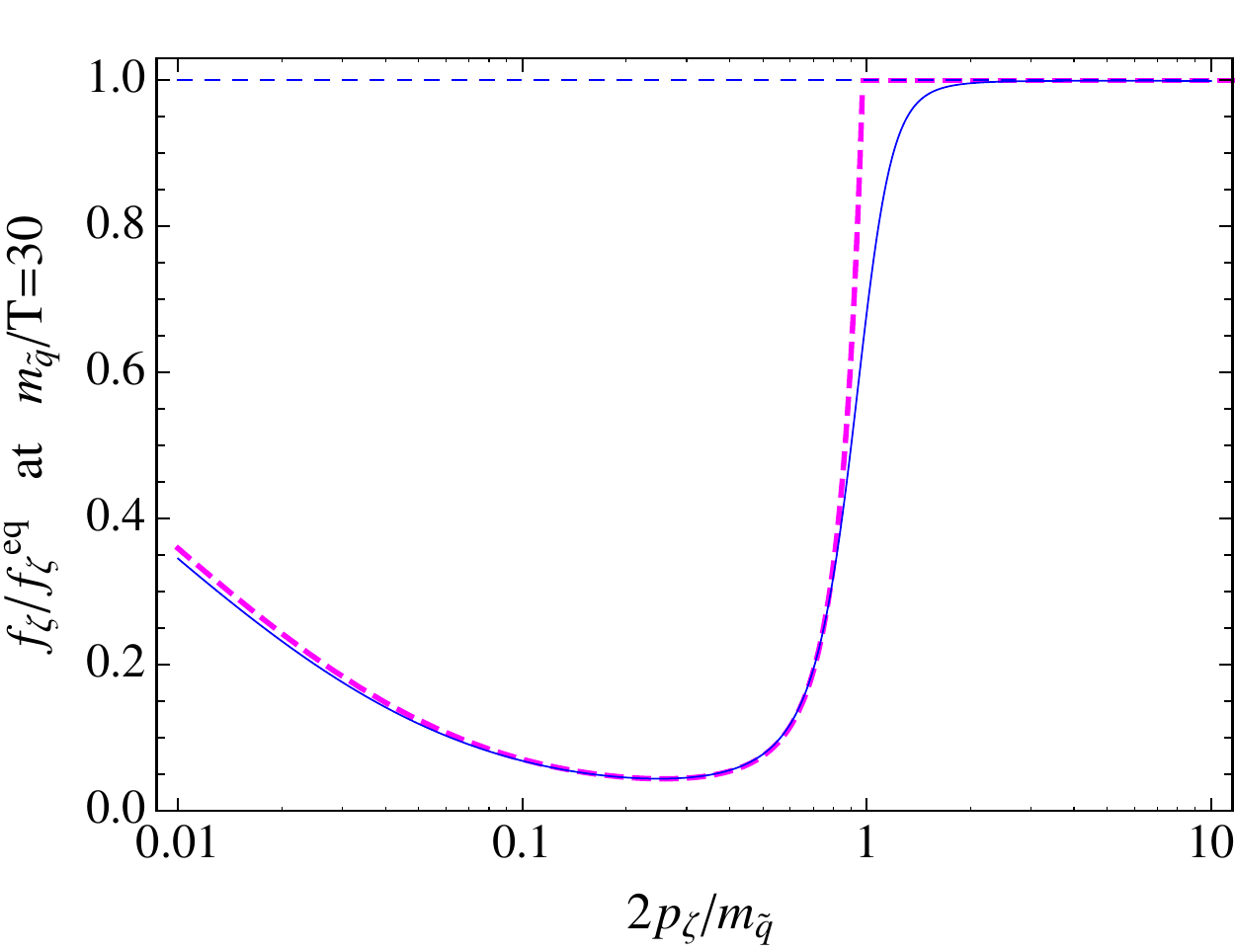}
\caption{Left: Effective ratio between the production rate and the Hubble parameter, as a function of the goldstino momentum and for different temperatures. Interactions are in chemical equilibrium when $R\z\gg1$, while the low-momentum region is frozen-out. Right: Ratio between the goldstino distribution function and the equilibrium distribution function, at $\mtq/T=30$. The continuous line comes from the numerical integration of the Boltzmann equation, while the magenta dashed line is the approximate analytical solution of Eq.~\eqref{fzeta}.}
\label{figRzeta}
\end{center}
\end{figure}

This behavior can be understood explicitly by looking at the RHS of the Boltzmann equation for the distribution function, Eq. \eqref{BoltzEQ}. We can estimate the effective ratio between the production rate and the expansion rate as
\beq\label{Rzeta}
%\frac{\Gamma_{\tilde q\to \zeta q} m_{\tilde q} T}{ H p_\zeta^2}
R_\zeta(p, a)\equiv  \sum_{\text{1 gen.}}\frac{\Gamma_{\tilde q\to \zeta q} \mtq T^2}{H p^3} \exp\left(-\frac{\mtq^2}{4 T p}\right).
%\sim \frac{\Delta f_\zeta|_{\rm collision}}{\Delta f_\zeta^{\rm eq}}
\eeq
For a given scale factor $a$ (corresponding to a given temperature $T(a)$), $R_\zeta(p, a)$ changes with the momentum; in particular, if $R_\zeta(p, a_0)\ll1$, the goldstinos with momentum $p$ are decoupled, while if  $R_\zeta(p, a_0)\gg1$ they are in equilibrium. In Fig. \ref{figRzeta}, we plot $R\z$ as a function of $p$ for different temperatures, fixing $\mtq=1\tev,\ F\z=(100\tev)^2,\ T_R=20\gev$. For example, at $T=\mtq/5=200\gev$ all goldstinos with momentum $p\gtrsim 50\gev$ %10^{-2}(\mtq/2)$
 are in thermal equilibrium, while at $T=\mtq/10 = 100\, {\rm GeV}
 %\,(\mtq/20 = 50\, {\rm GeV})
 $ only the goldstinos with momentum $p\gtrsim% 0.3 (\mtq/2)
 150\gev=1.5 T%\,(300\gev=6 T)
$ are, with the goldstino whose momentum is smaller than the temperature all decoupled. % low-momentum goldstinos frozen-out. %Because the number density is proportional to $\int p^3f\z(p)$
Thus at $T\approx \mtq/10$ the result is an earlier departure of the number density from its equilibrium value compared to that of assuming kinetic equilibrium, Eq.~\eqref{Tfo} %, where $f^{\rm eq}\sim e^{-p/T_0}$
(this earlier departure can also be seen in Fig.~\ref{decoupling_massive} in the Appendix).

Because $R_\zeta$ becomes very small once it is below one, we can solve the Boltzmann equation with zero RHS at low momentum,
\beq
&\frac{d f_\zeta(p, a)}{d\ln a} - \frac{ d f_\zeta(p, a)}{d\ln p} = 0 \quad &{\rm for}\quad p < p_{f.o.}(a),
\eeq
with boundary condition $f_\zeta(p,a) = e^{- p/T(a)}$ for $p \geq p_{f.o.}(a)$. Here  $p_{f.o.}(a)$ is the freeze-out momentum at a given temperature $T(a)$, defined by $R_\zeta( p_{f.o.}(a), a) = 1$:
\beq
&p_{f.o.}(a)=\frac{(k\z \mtq)^2}{T(a)},
\eeq
\begin{widetext}
\bea
%\label{kappazeta}
\kappa_\zeta = 0.13 \left(\frac{62.3}{62.3 + 4\ln\frac{T}{50\,{\rm GeV}} + 8 \ln\frac{(100\,{\rm TeV})^2}{F_\zeta} + 
2 \ln\frac{85}{g_*}  + 8\ln\frac{T_{\rm R}}{20\,{\rm GeV}} 
+ 24 \ln\frac{0.13}{\kappa_\zeta}}\right)^{1/2}.
\eea
\end{widetext}
Finally, the solution to the Boltzmann equation is 
\beq%\label{fzetaSOL}
&f_\zeta (p ,a) =
\left\{\begin{array}{ll}
\exp\left[ - \left(p/T_\zeta (a) \right)^{6/11}\right],	& p<p_{f.o.}(a)\\
\exp[-p/T(a)],		&p>p_{f.o.}(a)
\end{array}\right.
\label{fzeta}
\\
&T_\zeta(a) 
%= \left(\frac{T_i}{\kappa_\zeta m_{\tilde q}}\right)^{5/3}\left(\frac{a_i T_i}{a}\right)
= \left(\frac{T_R}{\kappa_\zeta m_{\tilde q}}\right)^{5/3}
\left(\frac{a_R T_R}{a}\right),
\eeq

To illustrate this better, in Fig. \ref{figRzeta}  we show the ratio between the resulting goldstino distribution function and the equilibrium distribution function, $f\z\propto e^{-p/T}$, for a given temperature $T=\mtq/30 \simeq 33\,{\rm GeV}$. In blue, we show the results from numerical integration of the Boltzmann equation; in dashed, the analytical expression for the distribution function, Eq.~\eqref{fzeta}, is shown, in good agreement with the numerical results. It is seen that the goldstinos at high momentum are in equilibrium, while the low-momentum ones are suppressed. %As a result, the final number density will also be reduced with respect to the result in Eq. \eqref{nzetaeq}. 

\begin{figure}[t]
\begin{center}
\includegraphics[height=0.22\textheight]{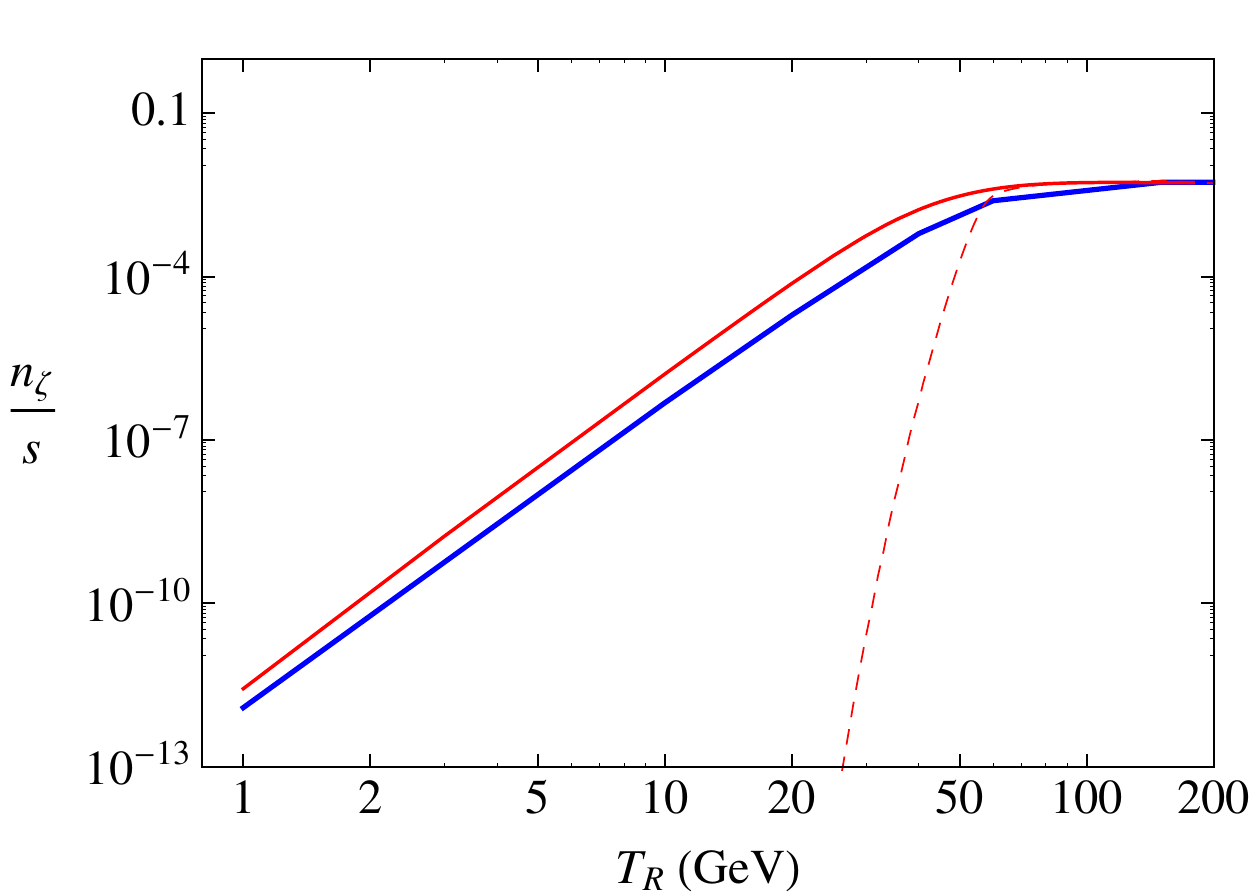}\hfill
\caption{Late-time goldstino yield as a function of the reheating temperature $T_R$, given $F\z=(100\tev)^2$ and $\mtq=1\tev$. The blue line represents the numerical results, which coincide with the analytical result of Eq.~\eqref{nzetaFO}, while the red line is the result that one would have found if kinetic equilibrium was assumed. The dashed red line is the yield found neglecting the matter-dominated era.
}
\label{zetayield}
\end{center}
\end{figure}

The late-time yield at low temperatures $T\lesssim T_R$, is given by
\bea\label{nzetaFO}
\left(\frac{n_\zeta}{s}\right)_{FO} = 6.8\times 10^{-7}
\left(\frac{85}{g_*}\right)\left(\frac{T_R}{10\,{\rm GeV}}\right)^5
\left(\frac{130\,{\rm GeV}}{\kappa_\zeta\mtq}\right)^5. \quad
\eea
This is shown in Fig. \ref{zetayield}, where we also show the result following from the kinetic equilibrium assumption, Eq.~\eqref{nzetaeq}, and the abundance found without considering the matter-dominated epoch, that is the case $T_R=T_{MAX}$, in which the Boltzmann suppression of the superpartner number density results in a negligible goldstino abundance at low reheating temperatures. Comparing the blue and magenta lines, we can conclude that a naive treatment of the Boltzmann equation overestimates the abundance by a factor of about 3.

As a reminder, these results were found in the limit of small $m\z$, but they do not change much if the goldstino mass is sizeable. Even for $m\z=100\gev$, the final yield changes only by about 10\%. In the Appendix, we show the full Boltzmann equation for the massive case, as well as numerical results for $m\z$ up to 200\gev\ (see Fig. \ref{decoupling_massive}).

Finally, as the distribution function deviates from the kinetic equilibrium case, one can also consider if the goldstinos produced would form colder or warmer dark matter, when compared to the case in which the particles are in kinetic equilibrium. 
The average momentum for the goldstino at $T\lesssim T_R$ can be evaluated as
\bea
\langle p\rangle \simeq 26 T_\zeta(a) = 
0.36\, T\left(\frac{g_*(T)}{g_*}\right)^{1/3}  \left(\frac{T_{R}}{10\, {\rm GeV}}\right)^{5/3}
 \left(\frac{130\, {\rm GeV}}{ \kappa_\zeta \mtq}\right)^{5/3}.
\eea
This should be compared to the thermal averaged value $\langle p \rangle_T\simeq 3T$. For low reheating temperatures, goldstinos are colder than the background temperature. This is due to two competing effects: at first, they are produced at a higher momentum, $p\simeq \mtq/2$, after which they are redshifted %by the universe expansion
 between production and reheating. For low $T_R$, the second effect is dominant.

\subsection{Non-Thermal Production}

Goldstinos can also be produced non-thermally, for example by direct moduli/inflaton decays, or by squark (or other lightest  WIMP particles in the MSSM) decays after freeze-out, for which $n\z/s=n_{\tilde q}/s$. The former is a model-dependent effect and it can be sizable or not. For what concerns the latter, it can be important  for large $F$-terms:
first of all the life-time of squark (with  $\mtq=1\tev$) 
has to be short enough to decay before BBN, implying  $F_\zeta\lesssim (10^5{\rm TeV})^2$ \cite{Cheung:2010mc}.  
Then, if $F_\zeta\gtrsim (5\times 10^4 {\rm TeV})^2$, 
the life-time of squark is long enough to decay after squark freeze-out by pair annihilation.  
In this case, the resulting energy density of the goldstinos is given by the non-thermal contribution
\beq
\Omega\z h^2=\frac{m\z}{\mtq}\Omega_{\tilde q} h^2.
\eeq
Since the reheating temperature is lower than the freeze-out temperature of squark annihilation, $T_{fr}^{\tilde q\tilde q*}$,   $\Omega_{\tilde q}h^2$ is also diluted by a factor of  $(T_R/T^{\tilde q\tilde q*}_{fr})^3$ compared to 
usual freeze-out abundance with $T_R\to\infty$ \cite{Giudice:2000ex}. 
Because of the substantial model-dependence in the non-thermal goldstino abundance, the results in Eqs.~\eqref{nzetaFI}-(\ref{nzetaFO}) should be considered as conservative results, as it is always possible to produce more goldstinos by introducing non-thermal processes.

\section{Late-time implications}\label{dmbbn}

The goldstinos produced in the early matter-dominated era are generally lighter than any other superpartner, except the gravitino. As such, they can provide a meta-stable dark matter candidate, even for very low reheating temperatures, $T_R\sim 1\gev$. The lifetime for the decay to a gravitino, $\zeta\to\psi_{3/2}\psi_{SM}\bar\psi_{SM}$ was computed in Ref.~\cite{Cheung:2010mc} as:
\beq\label{goldstinilifetime}
\tau\z\approx10^{22}\sec	\frud{F\z}{(100\tev)^2}^2\frd{100\gev}{m\z}^7.
\eeq
Although this is typically larger than the age of the universe, $t_0\simeq10^{17}\sec$, indirect detection limits on decaying dark matter are more stringent, with lower limits $\tau_{DM}\gtrsim \mathcal{O}(10^{26}-10^{27})\sec$ for dark matter decaying to quark-antiquark pairs \cite{Bertone:2007aw,Massari:2015xea,Giesen:2015ufa}. As a crude estimate, we take  the same order of magnitude, $\tau\z^{min}\approx 10^{26}\sec$ for the limits on the decaying goldstino.
These can be avoided with goldstini lighter than 100\gev, or larger $F\z$.

\subsection{Assuming \rpa}

\begin{figure}[t]
\begin{center}
\includegraphics[height=0.25\textheight]{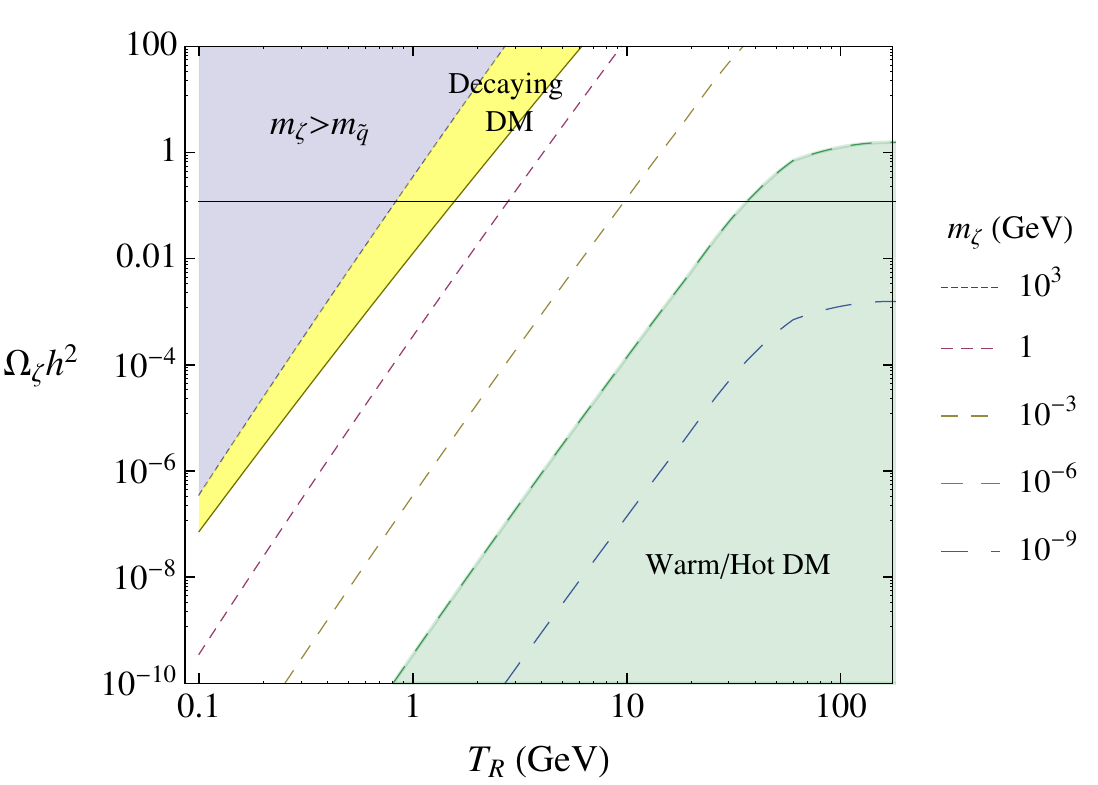}
\caption{Late-time energy density of goldstinos, with small $F$-term $F\z=(100\tev)^2$ and $\mtq=1\tev$, for different values of the goldstino mass. For $m\z<1 \,\rm keV$ the goldstino is warm or hot dark matter. For $m\z$ in the range $100\gev-1\tev$  (yellow region) the goldstino lifetime, Eq.~\eqref{goldstinilifetime}, is too short, $\tau\lesssim 10^{26}\sec$, and is excluded by DM indirect detection constraints.
The horizontal black line marks the observed value of the DM abundance, $\Omega_{DM}h^2=0.12$\,.
}
\label{omegazeta}
\end{center}
\end{figure}

If \rpa\, is conserved, the goldstino is effectively stable in most of the parameter space. For small $F$-term (corresponding to goldstinos produced before freeze-out), the present dark matter density is
\beq
(\Omega_\zeta h^2)_{FO} = 0.19 \left(\frac{m_\zeta}{1\,{\rm MeV}}\right) \frud{85}{g_*}
\left(\frac{T_{\rm R}}{10\,{\rm GeV}}\right)^5
\left(\frac{130\,{\rm GeV}}{k\z\mtq}\right)^5.
\eeq
The allowed region is shown in Fig.~\ref{omegazeta}, and spans the range $0.5\gev\lesssim T_R\lesssim30\gev$. The result is only logarithmically dependent on the increase of $F\z$, until $F\z=(500\tev)^2$, for which $T_{f.o.}>\mtq/10$. 
For larger $F$-term, the relevant process is freeze-in, and the dark matter abundance can be evaluated from Eq.~\eqref{nzetaFI}:
\beq
(\Omega_\zeta h^2)_{FI} =0.11 \frud{m\z}{2\,\rm MeV}
\left(\frac{85}{g_*} \right)^{3/2} 
\left(\frac{T_{\rm R}}{10\, {\rm GeV}}\right)^7 \left(\frac{(500\, {\rm TeV})^2}{F_\zeta}\right)^2 
\left(\frac{1\,\rm TeV}{m_{\tilde q}}\right)^4 .
\eeq

To summarize this section, the observed dark matter abundance can easily be produced in a matter-dominated era, at temperatures above a low reheating temperature $T_R$.

\subsection{Assuming \rpv}

Another well-motivated possibility is that of \rpv. In the general case, baryon and lepton number would be violated and the proton would be unstable. Nevertheless, if lepton or baryon number were to be independently conserved on their own, proton stability would be achieved accidentally. In the following, we will discuss the case in which baryon number is violated while lepton number is conserved (the other case will have a similar phenomenology). 
This is also interesting because baryonic \rpv\ can account for the matter-antimatter asymmetry of the universe with low reheating temperatures, while other scenarios (such as leptogenesis) require higher temperatures. The baryonic RPV operator in the superpotential is
\beq
W_{BRPV}=\frac{\l_{ijk}}{2}u_i^cd_j^cd_k^c+h.c.\,,
\eeq
where the contraction of the color indices with an $\epsilon^{abc}$ tensor is understood and as a consequence $j\neq k$.

In low-scale gauge mediation with a single SUSY breaking sector, the gravitino is very light and there are still proton decay channels of the type $p\to K\psi_{3/2}$, mediated by RPV interactions. The resulting limits are very stringent and were discussed in \cite{Choi:1996nk,Choi:1998ak}. 
In the case of multiple goldstini, the proton could potentially decay to any goldstino lighter than 1\gev. This is particularly dangerous when  $F\z$ is small, independently of the goldstino mass: for example, the weakest limit is
\beq
\lambda''_{323}< %5\times 10^{-8} \fru{\mtq}{300\gev}^2\fru{\mtr}{1 eV}
%5.5\times 10^{-7} \fru{\mtq}{1\tev}^2\fru{\mtr}{1 eV}=
1.31\times 10^{-6} \fru{\mtq}{1\tev}^2\frac{F\z}{(100 \tev)^2}\,.
\eeq
To avoid this constraint, we require that $m\z>m_p$; 
%the gravitino mass would also need to be large, corresponding to $F_{eff}\simeq (10^{5}\tev)^{2}$.
%this is natural in the case of multiple low-scale SUSY breaking sectors \cite{Argurio:2011hs}, or in the case of one gauge-mediated and one gravity-mediated sector \cite{Cheung:2010mc}. The gravitino mass would be small in the former case, $\mtr\ll m\z$ (a large $F_{eff}\simeq (10^6\tev)^2$ would be needed to avoid proton decay to gravitino), while the two masses would be comparable in the latter, $m\z\simeq2\mtr$. 
however, now it is the goldstino which is unstable, as the decay channel $\zeta\to u_id_jd_k$ is open. The lifetime is
\beq\label{tauzeta}
\tau_\zeta= & 1.57\times 10^{3}\sec
\left(\frac{1}{\lambda''_{ijk}}\right)^2\left(\frac{10\, {\rm GeV}}{m_{\zeta}}\right)^9 
\left(\frac{m_{\tilde q}}{1 {\rm TeV}}\right)^4\left(\frac{F_\zeta}{(100\,{\rm TeV})^2} \right)^2 \,,
\eeq
where $\l_{ijk}$ is the largest RPV coupling for which the decay is kinematically accessible.
%depends on the goldstino mass (for $m\z>m_b+m_c+m_s$, the relevant operator is $\l_{223}c^cb^cs^c$, while for $m\z>m_t$ the decays through the operators $\l_{3ij}td_id_j$)
As the goldstino mass naturally lies in the interval $1-100$\gev, the top quark is not accessible and the most relevant operator with few constraints from flavor physics is $\l_{223}c^cb^cs^c$ \cite{Barbier:2004ez}.

The lifetime \eqref{tauzeta} of the goldstino naturally falls in a range that is probed by Big Bang Nucleosynthesis: if a large amount of energy is injected during the thermal plasma during BBN, the primordial abundance of light elements is changed and would go against observations. In particular, the case of hadronic decays was studied in great details in Refs. \cite{Kawasaki:2004yh,Kawasaki:2004qu,Jedamzik:2004er,Jedamzik:2006xz}. In the following, we will use the results of Ref. \cite{Kawasaki:2004qu}, where limits on the abundance $M_X Y_X$ of a decaying particle $X$ were set in the lifetime range $10^{-2}\sec<\tau_X<10^{12}\sec$, for different masses $M_X=100\gev,1\tev,10\tev$. 

As we are also interested in particles with lighter masses, we need to extrapolate their results to $M_X=10\gev$ and below. Therefore, we will shortly review the source of the limits. For short lifetimes ($\tau<10^2\sec$), the mesons and nucleons produced by $X$ thermalize quickly, and the main consequence of the decay is the increase of the neutron-to-proton ratio, $n/p$, resulting in larger abundances of D and \hef. For longer lifetimes, mesons decay and primary protons and neutrons scatter inelastically off of the background nuclei, generating hadronic showers, dissociating \hef \ and producing D, T, \het, which also result in higher amounts of \lis, \lisv. At $\tau>10^7\sec$, the neutrons decay away and only protons are left, with a smaller effect on \hef-dissociation. On the other hand, electromagnetic decay products ($\gamma,e^+,e^-$) are thermalized by processes such as $\gamma+\gamma_{BG}\to e^++e^-$ if their energy is above the threshold $E_{th}={m_e^2}/{22T}$ \cite{Kawasaki:1994af}; one should compare this threshold to the binding energy of D and \hef, respectively 2.2 and 28.3\mev: if it is higher, non-thermalized photons will dissociate deuterium and helium. As the bath temperature decreases with time, photo-dissociation of D and \hef\ is active for $\tau>10^4\sec$ and $\tau>10^6\sec$, respectively.

\begin{figure}[t]
\begin{center}
\includegraphics[height=0.3\textheight]{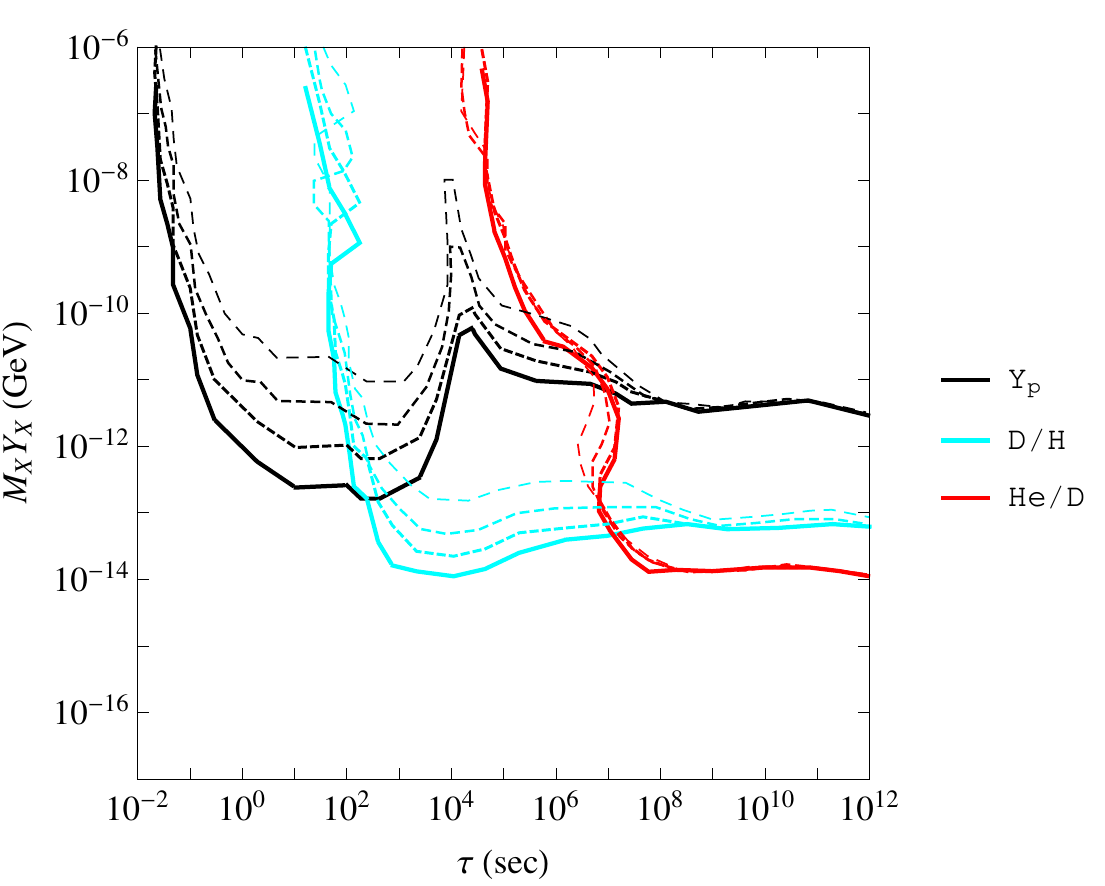}\hfill 
\includegraphics[height=0.3\textheight]{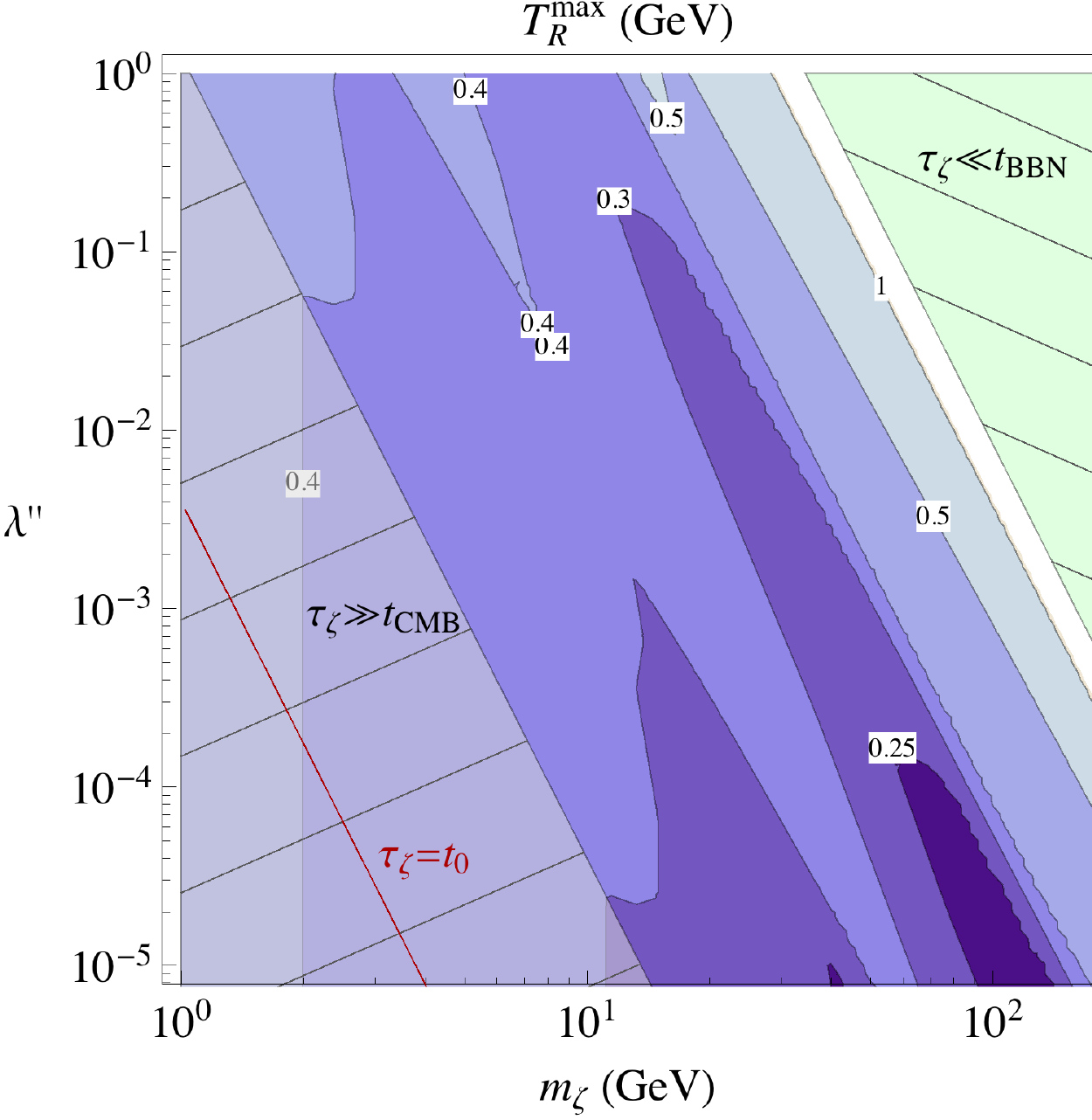}
\caption{Left: Upper limits on the energy density of a hadronically decaying particle $X$, with, from bottom to top, $M_X=10\gev,100\gev,1\tev,10\tev$, evaluated from Ref. \cite{Kawasaki:2004qu}. Right:
Maximum reheating temperature allowed by BBN constraints while varying $\l_{ijk}$ and $m\z$. The other parameters are fixed as $F\z=(100\tev)^2$ and $\mtq=1\tev$.}
\label{figbbn}
\end{center}
\end{figure}

We simulate the goldstino decay with Pythia 8.2 \cite{Sjostrand:2014zea} and get the total number of charged particles and EM energy per $\zeta$ decay, for different values of $m\z$. We then translate the results of Ref.~\cite{Kawasaki:2004qu} to lower masses. For the sake of simplicity, we only use the dominant constraints, that is primordial helium abundance ($Y_p$), deuterium to hydrogen ratio (D/H) and helium-4 to deuterium ratio (\hef/D).
Our results are shown in Fig.~\ref{figbbn}, where for comparison we also show the constraints of Ref.~\cite{Kawasaki:2004qu}.

For most of the lifetime range, the limits on $m\z Y\z$ are of  order $10^{-12}-10^{-14}\gev$. Comparing this to the yields found in Eqs.~\eqref{nzetaFI} and \eqref{nzetaFO}, only very low reheating temperatures are allowed. In Fig.~\ref{figbbn}, we show contours of the maximum allowed reheating temperatures in the $m\z-\l_{ijk}$ plane: apart from a small shaded region in the top right corner, for which $\tau\z<10^{-2}\sec$, the upper limit on the reheating temperature is of order 0.5\gev. In the lower left corner, we shaded the area in which $\tau\z\gg10^{12}\sec$ and goldstino decays do not interfere with BBN: here they would be constrained by diffuse $X$-ray and $\gamma$-ray emissions, and the limits on $m\z Y\z$ are typically more strict than the BBN ones. As this discussion is beyond the scope of the present paper, we only display the BBN limits as a conservative bound.

In the presence of \rpv, there is one more channel for the production of goldstini that had not been analyzed so far in the literature: the scattering $qq\to q\zeta$ does not require any on-shell superpartners, in contrast with the \rpcing processes discussed above.
Using the interaction in Eq.~(\ref{RPVint}), the freeze-out temperature for the scattering is
\beq
T_{f.o.}^{\rm RPV} = 70\, {\rm GeV} \left(\frac{g_*}{85}\right)^{1/10}\left(\frac{F_\zeta}{(100\,{\rm TeV})^2}\right)^{2/5}
\left(\frac{m_{\tilde q}}{{\rm TeV}}\right)^{4/5} 
\left(\frac{1}{\lambda''}\right)^{2/5}\left(\frac{10\, {\rm GeV}}{T_{\rm R}}\right)^{2/5}.
\eeq
The goldstino distribution is the same as that in kinetic equilibrium as long as they are in chemical equilibrium, and once we solve the relevant Boltzmann equation we find the yield,
\beq
\left(\frac{n_\zeta}{s}\right)_{\rm RPV} =&
%3.2\times 10^{-7}\left(\frac{85}{g_*}\right) 
%\left(\frac{T_{\rm R}}{10\,{\rm GeV}}\right)^5\left(\frac{70\,{\rm GeV}}{T_{f.o.}^{\rm RPV}}\right)^5
%\\ 
%&\qquad
%+\,
\, 7.0 \times 10^{-7}|\lambda''|^2  \ln\frac{T_{f.o.}^{\rm RPV}}{T_{\rm R}} 
\left(\frac{85}{g_*} \right)^{3/2} 
\left(\frac{T_{\rm R}}{10\, {\rm GeV}}\right)^7 \left(\frac{(100\, {\rm TeV})^2}{F_\zeta}\right)^2 
\left(\frac{\rm TeV}{m_{\tilde q}}\right)^4 \nn
\\
&+\,
4.7 \times 10^{-8}|\lambda''|^2 \left(\frac{85}{g_*} \right)^{3/2} 
\left(\frac{T_{\rm R}}{10\, {\rm GeV}}\right)^7 \left(\frac{(100\, {\rm TeV})^2}{F_\zeta}\right)^2 
\left(\frac{\rm TeV}{m_{\tilde q}}\right)^4. 
\eeq
Here the first line shows the production during the matter-dominated era and the second line is the production after reheating; in general, the former dominates over the latter, and the \rpcing freeze-out contribution from Eq. \eqref{nzetaFO} is larger than both.
For the RPV production rate to be sizable, the only option would be $T_R=T_{MAX}\ll\mt$, for which there is no early matter-dominated era. In this case, only the last line in the equation above contributes to  goldstino production. 
There are still strong limits from BBN, and the maximum reheating temperature is of order $1-10\gev$, with smaller RPV couplings allowing slightly larger reheating temperatures ($T_R^{max}\simeq20\gev$ for $\l=10^{-5}$).

The limits on the reheating temperature cited so far corresponded to $F\z=(100\tev)^2$, with goldstinos generated at  freeze-out. The dependence of $T_R^{max}$ on larger $F$-terms is first logarithmic, until $F\z\simeq (500\tev)^2$, and then scales as $T_R^{max}\propto (F\z/(500\tev)^2)^{2/7}$ when the freeze-in contribution \eqref{nzetaFI} becomes dominant. For example, $F\z=(10^6\tev)^2$ corresponds to $T_R^{max}\simeq 100\gev$ for $m\z=10\gev$ and $\l_{ijk}=1$.

We end this section by summarizing the strong bounds on the reheating temperature in the case in which \rpa\ is violated.
%In most of the parameter space, reheating temperatures above $1-100\gev$ are excluded.
%Apart from a region with large $m\z$ and large RPV coupling, the maximum reheating temperature is less than about 1\gev.
For the freeze-out case, the maximum reheating temperature allowed by BBN constraints is of order 1\gev.

Some implications for what concerns baryogenesis can be drawn: because any baryon asymmetry produced at higher temperatures in the matter-dominated era will be diluted away, the baryon asymmetry should be generated  at temperatures between $T_R$ and $T_{BBN}\sim10\mev$. This is possible in the LSP baryogenesis scenario of Ref.~\cite{Monteux:2014hua} if the goldstino decays before BBN (in the upper right corner of Fig.~\ref{figbbn}): for example, with  parameters  chosen as $F\z=(100\tev)^2$, $\l_{ijk}\simeq 0.1$, the goldstino abundance in Eq.~\eqref{nzetaFO} can be large enough for baryogenesis if $m\z\simeq T_R\simeq 50\gev$.

%a viable option would be for the baryon asymmetry to be directly generated in goldstino decays, at temperatures between $T_R$ and $T_{BBN}\sim10\mev$. This is possible in the LSP baryogenesis scenario of Ref.~\cite{Monteux:2014hua}: for example, a goldstino with mass $m\z=10\gev$ could generate the observed baryon asymmetry if $\epsilon$, the baryon asymmetry per particle decay, is of order $10^{-5}$.

Even though we have taken $\mtq=1\tev$ as a benchmark point, it is worth noting that current LHC limits on RPV squarks are less stringent: for a light top squark decaying to three quarks, the CMS collaboration excludes masses up to 350-385\gev\, in Ref. \cite{Khachatryan:2014lpa}. Gluino limits are stronger: in Ref. \cite{Chatrchyan:2013fea}, CMS excludes gluinos in the decay channel $\tilde g\to t b s$ up to 900\gev, while in the same channel ATLAS excludes  gluinos up to 874\gev, with limits of order 800\gev\, for different flavor composition of the final states.
Thus, squarks are still allowed to be lighter than in our benchmark point.

\section{Conclusions}\label{end}
In this work, we have discussed thermal production of goldstinos as an example of super-weakly interacting particles during an early matter-dominated era ending at reheating. This is important when the reheating temperature is low, in the GeV range or below, as particle production through %some other
an heavier sector (superparticles) occurs only at higher temperatures that were not achieved during radiation domination.

We have analyzed in detail the production of an uneaten goldstino by solving the Boltzmann equation for its momentum distribution function, and revisited the cosmological implications. When the goldstino is stable enough, thermal production can provides the correct dark matter density even for reheating temperatures as small as $1\, { \rm GeV}$.
%for a reheating temperature well below the superpartners mass scale, and found that it can be the dark matter . 
If \rpa\ is violated, the goldstino has to be heavier than the proton and is meta-stable, with a lifetime range that naturally interferes with BBN. In this case, reheating temperatures higher than 1\gev\ are excluded for almost all of the small $F\z$ parameter space.

Such low reheating temperatures suggest a low scale of inflation and/or introducing certain symmetries to prevent coupling between the inflaton and visible fields. For example, if the inflaton decay rate to the MSSM is Planck-suppressed, $\Gamma_I={m_I^3}/{M_P^2}$, the reheating temperature is
%\beq
$T_R \simeq 1\gev ({m_I}/{10^3\tev})^{3/2}$.
%\eeq
In this case we can also find an upper bound on $T_{MAX}$ from its definition in Eq.~\eqref{Tmax},
%\beq
$T_{MAX}\lesssim 10^3\tev (T_R/\gev)^{2/3}$.
%\eeq 
Since the non-thermal production is 
proportional to $T_R/m_I\propto T_R^{1/3}$, it could be more important at low reheating temperature, and full analysis considering a specific inflation model is needed. 
In this paper, we presented thermal production of goldstino as model independent contributions. 
Our results are conservative bounds on the abundance, because
the non-thermal productions of SWIMP are just additive quantities. 
%For low reheating temperatures, the non-thermal production from moduli/inflaton decay can typically be large. For example, the yield would be of order $n/s\sim 10^{-6}({m_I}/{10^3\tev})^{1/2}$.

%OTHERS? LHC? X($\gamma$-ray)?  

We showed that the goldstino is one of many good examples in which a momentum dependent process gives a sizable difference compared to that assuming thermal distribution of the momentum, even if they are produced from thermal bath.
Such effects are also discussed in Ref.~\cite{Shi:1998km} for the production of sterile neutrino dark matter, 
and in Ref.~\cite{Basboll:2006yx} for leptogenesis from heavy Majorana neutrino decays. 
Since the period of kinetic decoupling and production mechanism are also important to
the evolution of density perturbation of dark matter, the study of the full Boltzmann equations in the perturbed spacetime can give observable consequences for small scale structures. 

\begin{acknowledgments}
We would like to thank Matt Buckley, David Shih, Scott Thomas and Stefano Profumo for useful discussions about this subject. A.M. and C.S.S. are supported in part by DOE grants DOE-SC0010008, 
DOE-ARRA-SC0003883, and DOE-DE-SC0007897.

\end{acknowledgments}

\appendix

\section{Boltzmann Equation for Massive Goldstino}\label{appendix}
In this section we derive the Boltzmann equation for the momentum distribution function of $\zeta$, $f_\zeta({\bf p})$, keeping the dependence on a non-zero mass  $m_\zeta$. 
The dominant source of production is sfermion decays, $\phi \to\zeta+ \psi$, when the temperature is lower that the sfermion mass $\tilde m_{\phi}$, and the contribution from elastic scattering is ignored. 
As we did in the rest of this work, we will consider the Boltzmann equation and distribution functions in the classical limit, i.e. assuming 
$f_\chi({\bf p})\lesssim1$, and $f_\chi^{\rm eq} ({\bf p})\simeq e^{-E_\chi/T}$, where $E_\chi = \sqrt{m_\chi^2 + {\bf p}^2}$ 
for particle $\chi$.
 The corresponding Boltzmann equation in the limit of massless quarks is 
\bea
\frac{d f_\zeta}{d t} &=&
\frac{\partial f_\zeta}{\partial t}  - H p \frac{\partial f_\zeta}{\partial p}= C[f_\zeta]
\nonumber\\
&=& \frac{g_{\phi} g_\psi}{2 E_\zeta} \int \frac{d^3{\bf p}_\phi}{(2\pi)^3 2 E_{\phi}} 
\frac{d^3 {\bf p}_\psi}{(2\pi)^3  2 p_\psi} (2\pi)^4 \delta^{(4)}(p^\mu_{\phi} - p^\mu_\psi - p^\mu) 
\left|{\cal M}_{\phi\to \zeta \psi}\right|^2 (f_{\phi} - f_\zeta f_\psi),
\eea where $p =|{\bf p}|$. 
Tree level $T$ symmetry ensures $|{\cal M}_{\phi\to \z \psi}|^2 = |{\cal M}_{\z \psi\to \phi}|^2$ at leading order.
Sfermions are in thermal equilibrium, which is maintained by the interactions with the background SM fields ($\phi+\phi^*\leftrightarrow A_\mu+A_\mu$, $\phi+ \psi \to \phi+ \psi$, $\ldots$) before such interactions are frozen.
After that, the distribution of sfermions is determined by the interaction with the goldstinos, and 
we need to solve the coupled Boltzmann equations. 
However, most of goldstinos are produced before the freeze-out of sfermion-SM interactions, so we can safely take 
$f_{\phi} ({\bf p}_{\phi}) = e^{- E_{\phi}/T}$.

Using the identity $f_{\phi}^{\rm eq} \delta^{(4)}(p_{\phi}^\mu - p_\psi^\mu - p^\mu) 
= f_\zeta^{\rm eq} f_\psi^{\rm eq}\delta^{(4)}(p_{\phi}^\mu - p_\psi^\mu - p^\mu)$ to represent $f_\psi$, after integrating over ${\bf p}_\psi$, we get
\bea
C[f_\zeta] &=& 
\left(1-\frac{f_\zeta}{f_\zeta^{\rm eq}}\right)
\frac{g_{\phi} g_\psi|{\cal M}|^2}{2E_\zeta}\int \frac{dp_{\phi}}{(8\pi)} f_{\phi}^{\rm eq}\, \frac{ p_{\phi}^2 }{E_{\phi}  p_\psi^*} 
\int d\cos\theta_{\phi\zeta}\delta(E_{\phi} - E_\zeta - p_\psi^*)
\nonumber\\
&=&
\left(1-\frac{f_\zeta}{f_\zeta^{\rm eq}}\right)
\frac{g_{\phi} g_\psi|{\cal M}|^2}{16\pi E_\zeta}\int_{D[p]} dp_{\phi}\,  \frac{ p_{\phi}^2 }{E_{\phi}  p_\psi^*} 
 \frac{ e^{- E_{\phi}/T}}{|dp^*_\psi/d\cos\theta_{\phi \zeta}|}\,,
\eea
where 
$
p^*_\psi =\sqrt{ p_{\phi}^2  + p^2 - 2 p_{\phi} p\cos\theta_{\phi \zeta}},
$
so $|d p^*_\psi/d\cos\theta_{\phi \zeta}|= p_{\phi} p/p^*_\psi $.
For given $p$, the integral domain $D[p]$ is  determined by 
$
E_{\phi} - E_\zeta = \sqrt{p_{\phi}^2 + p^2 - 2 p_{\phi} p \cos\theta_{\phi\zeta}}$ 
for $-1 \leq \cos\theta_{\phi\zeta}\leq 1$. 
Then, we find $p_{\phi}^- \leq p_{\phi}\leq p_{\phi}^+$, where 
\beq
p_{\phi}^{\mp} 
&= m_{\phi}^2 \frac{ (E_\zeta \mp p)}{2m_\zeta^2}  - \frac{(E_\zeta \pm p)}{2}.
\eeq
In terms of energy variable $E_{\phi}$, $E_{\phi}^- \leq E_{\phi} \leq E_{\phi}^+$, where 
\beq
E_{\phi}^\mp 
&= m_{\phi}^2\frac{( E_\zeta \mp p)}{2 m_\zeta^2} + \frac{(E_\zeta \pm p)}{2}.
\eeq 
In massless limit, $m_\zeta\to 0$, we obtain  
\bea
 \frac{m_{\phi}^2}{4p} + p < E_{\phi}, 
\eea which was used in Eq. (\ref{BoltzEQ}). 
It is straightforward to evaluate $C[f_\zeta]$ as 
\bea
C[f_\zeta] &=&
\left( 1- \frac{f_\zeta}{f_\zeta^{\rm eq}}\right) 
\frac{g_{\phi} g_q|{\cal M}|^2}{16\pi  E_\zeta p}
\int_{E_{\phi}^-}^{E_{\phi}^+} 
d E_{\phi}\,    e^{- E_{\phi}/T} \nonumber\\
&=&\left(1 -\frac{f_\zeta}{e^{- E_\zeta/T}}\right) 
\left(\frac{g_{\phi} \Gamma_{\phi\to\zeta q} m_{\phi} T}{g_\zeta E_\zeta p}\right)
\left[e^{-E_{\phi}^-/T}- e^{-E_{\phi}^+/T}\right].
\eea
Finally, the Boltzmann equation can be written as
\bea
\frac{\partial f_\zeta}{\partial \ln a } - \frac{\partial f_\zeta}{\partial \ln p}  &=&
\left(1 - e^{E_\zeta/T} f_\zeta\right)\left(\frac{\Gamma_{\phi
\to \zeta q} m_{\tilde q} T}{ H E_\zeta p}\right)
\left[e^{-E_{\phi}^-/T}- e^{-E_{\phi}^+/T}\right]
.\eea
This expression replaces Eq.~\eqref{BoltzEQ} in the case of non-negligible $m\z$.

\begin{figure}[tb]
\begin{center}
\includegraphics[height=0.22\textheight]{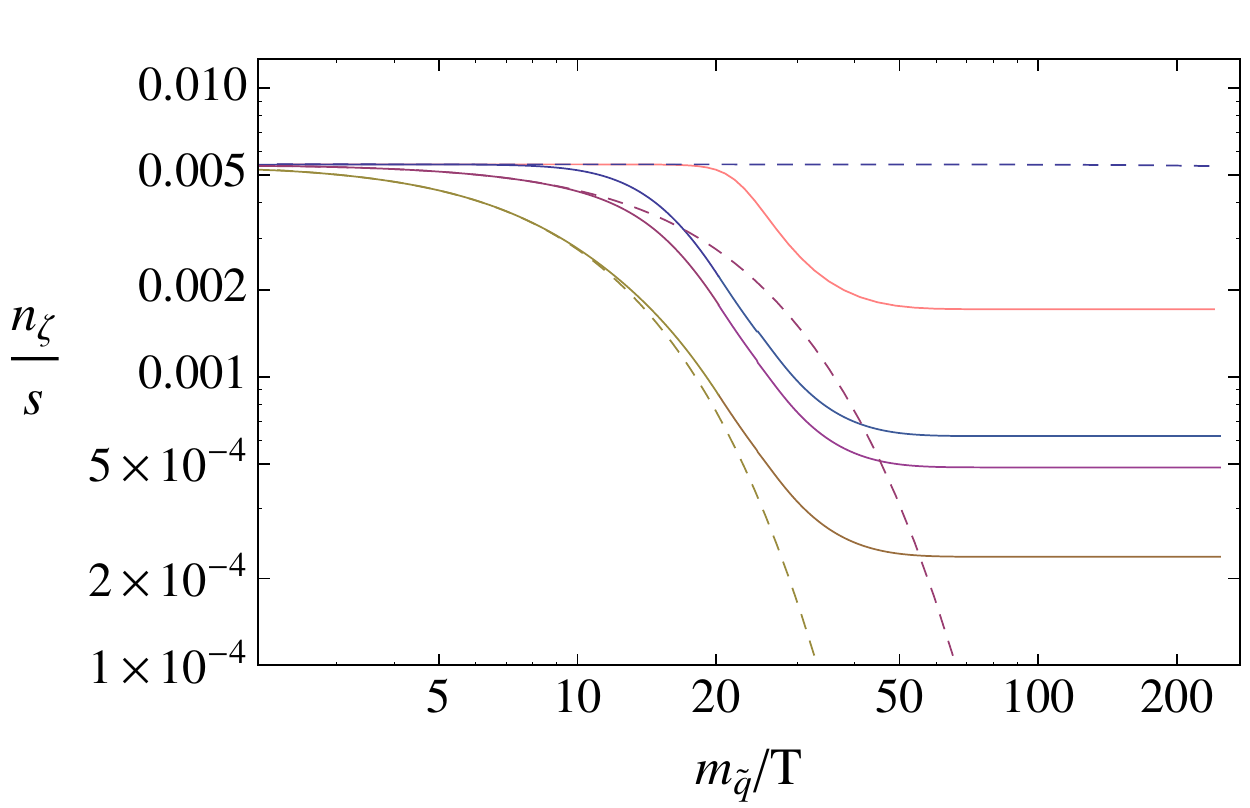}
\caption{Evolution of the goldstino number density for non-negligible $m\z$. Here we have fixed $\mtq=1\tev,\,F\z=(100\tev)^2$ and $T_R$=40\gev. From top to bottom, the continuous lines are the yield computed assuming kinetic equilibrium in the massless case, and the yields using the full Boltzmann equation, for three different values of the masses, $m\z=0\gev,100\gev,200\gev$. The dashed lines are equilibrium number densities for the same masses. %Note the earlier departure from chemical equilibrium with respect to the kinetic equilibrium assumption.
}
\label{decoupling_massive}
\end{center}
\end{figure}

For example, in Figure~\ref{decoupling_massive} we show the evolution of the goldstino number density for different values of the goldstino mass for a fixed $T_R=40\gev$. From top to bottom, with the continuous lines we show the yield computed assuming kinetic equilibrium and $m\z=0$, and the yields computed solving the above Boltzmann equation for three different values of the masses, $m\z=0\gev,100\gev,200\gev$. We find that the final yield of goldstinos is not changed much for $m\z\lesssim 100\gev$.  The dashed lines are the equilibrium number densities for the three different masses: we see that the effect of the masses is much smaller than what naively expected by looking at the equilibrium number density.

\bibliography{biblio}

\end{document}